%% file: main.tex
\documentclass[notitlepage,a4paper,superscriptaddress,longbibliography, 10pt,twocolumn,pra,footinbib]{revtex4-1}
\usepackage{graphicx}
\usepackage{amsfonts}
\usepackage{amssymb}
\usepackage{amsmath}
\usepackage{braket}
\usepackage{eucal}
\usepackage{diagbox}
\usepackage{microtype}

\usepackage{soul}

\usepackage{dsfont}

\usepackage{mathrsfs}
\usepackage{soul}
\usepackage[usenames, dvipsnames]{color}
\usepackage[colorlinks=true,citecolor=blue, urlcolor=black]{hyperref}

\usepackage{graphicx}
\usepackage{tikz}

\newcommand{\mean}[1]{\langle #1 \rangle}

\usepackage{xcolor}

\newcommand{\Tr}{\textnormal{Tr}}

\renewcommand{\Re}{\textnormal{Re}}
\renewcommand{\Im}{\textnormal{Im}}
\renewcommand{\deg}{\textnormal{deg}}

\newcommand{\SOS}{\textnormal{SOS}}

\newcommand{\normord}[1]{:\mathrel{#1}:}

\usepackage{amsthm}
\newtheorem{Theorem}{Theorem}
\newtheorem{Proposition}[Theorem]{Proposition}

\newcommand{\new}[1]{{\color{black} #1}}

\begin{document}
\title{Revealing hidden physical nonclassicality with nonnegative polynomials}

\author{Ties-A. Ohst}
\affiliation{Naturwissenschaftlich-Technische Fakultät, 
Universität Siegen, Walter-Flex-Straße 3, 57068 Siegen, Germany}
\email{ties-albrecht.ohst@uni-siegen.de}

\author{Benjamin Yadin}
\affiliation{Naturwissenschaftlich-Technische Fakultät, 
Universität Siegen, Walter-Flex-Straße 3, 57068 Siegen, Germany}

\author{Birte Ostermann}
\affiliation{Technische Universität Braunschweig, Institut für Analysis und Algebra, AG Algebra, Universitätsplatz 2, 38106 Braunschweig, Germany}

\author{Timo de Wolff}
\affiliation{Technische Universität Braunschweig, Institut für Analysis und Algebra, AG Algebra, Universitätsplatz 2, 38106 Braunschweig, Germany}

\author{Otfried G\"uhne}
\affiliation{Naturwissenschaftlich-Technische Fakultät, 
Universität Siegen, Walter-Flex-Straße 3, 57068 Siegen, Germany}

\author{Hai-Chau Nguyen}
\affiliation{Naturwissenschaftlich-Technische Fakultät, 
Universität Siegen, Walter-Flex-Straße 3, 57068 Siegen, Germany}
\email{chau.nguyen@uni-siegen.de}

\date{\today}

\begin{abstract}
Understanding quantum phenomena which go beyond classical concepts is a focus of modern quantum physics. Here, we show how the theory of nonnegative polynomials emerging around Hilbert's 17th problem, can be used
to optimally exploit data capturing the nonclassical nature of light. 
Specifically, we show that nonnegative polynomials can reveal 
nonclassicality in data even when it is hidden from standard 
detection methods up to now. 
Moreover, the abstract language of nonnegative polynomials also leads 
to a unified mathematical approach to nonclassicality for light and 
spin systems, allowing us to map methods for one to the other.
Conversely, the physical problems arising also inspire
several mathematical insights into characterisation of 
nonnegative polynomials.
\end{abstract}

\maketitle

\textit{Introduction.---}
The research on nonclassicality was initiated by the struggle to understand fundamental aspects of quantum mechanics. This is perhaps nowhere more 
vividly perceived than in the field of quantum optics.
Facing the challenge to refute the classical wave nature of light, 
reigning since Huygens and Fresnel, ingenious theoretical ideas and sophisticated experimental techniques have been developed~\cite{fox2006quantum,Gerry_Knight_2004}. 
Recently, the research on nonclassicality of light has been intensified, spurred by its promising applications in quantum technology, ranging 
from quantum cryptography~\cite{xu_secure_2020}, quantum imaging~\cite{moreau_imaging_2019}, to the 
celebrated enhancement of gravitational
detectors~\cite{the_ligo_scientific_collaboration_gravitational_2011,aasi_enhanced_2013a}.

Historically, the demonstration of photon antibunching and sub-poissonian counting statistics of photons marked the first realisations of optical phenomena unexplainable 
by the classical wave theory~\cite{kimble_photon_1977,short_observation_1983}. 
The demonstration of nonclassicality of light has since been generalised in different aspects, leading in particular to the development of the analysis of higher moments 
of photon counting statistics~\cite{PhysRevA.46.485} and eventually to the application of modern methods of sum of squares~\cite{PhysRevLett.94.153601} and moment matrices~\cite{PhysRevA.71.011802,PhysRevA.72.043808,Richter2003, PhysRevLett.89.283601}.

In contemporary quantum science, the concept of nonclassicality is investigated in a more general sense, which is in particular 
linked to quantum entanglement~\cite{Ghne2009, RevModPhys.81.865} and quantum correlations~\cite{RevModPhys.86.419,uola2020}. Of special interest 
is the concept of nonclassicality of systems of spin-$\frac12$
particles, which is motivated 
by that of light~\cite{PhysRevA.78.042112}. 
The concept of nonclassicality of spin systems bears 
an apparent similarity with entanglement theory~\cite{PhysRevA.78.042112}, leading to an independent development of methods for demonstration of nonclassicality for spin-$\frac12$ systems, also known as entanglement theory in the bosonic subspace~\cite{Wang2002, PhysRevA.67.022112, PhysRevLett.102.170503, Quesada2017, PhysRevA.94.042324, PhysRevA.87.012104, Qian2020, Eckert2002, Yu2016, PhysRevA.96.032312}.



In the present work, we show that detection of nonclassicality can go beyond what has been achieved by the previously mentioned methods. 
To this end, we consider observables whose expectation values for basic classical states are nonnegative polynomials in their parameters, such as the amplitude and phase of the light wave or the direction of the collective angular momentum of the spin system. 
It follows that their expectation values for all classical states are also nonnegative, and a violation indicates the nonclassicality.
Crucial to our finding is the existence of nonnegative polynomials that cannot be written as a sum of squares of polynomials, as first shown by David Hilbert in a seminal paper in 1888~\cite{Hilbert1888}.
The 17th problem in his 23 mathematical problems posed for the twentieth century concerns the question of whether such a nonnegative polynomial 
can always be represented as a sum of squares of \emph{rational functions}. 
This eventually led to the rich theory of nonnegative polynomials and 
real algebraic
geometry~\cite{Artin1927,Blekherman:Parrilo:Thomas,Lasserre2000GlobalOW, Lasserre:IntroductionPolynomialandSemiAlgebraicOptimization, Laurent:Survey, Theobald:Book, Delzell:Prestel:Survey,Nesterov:Nemirovski}. 
However, the relevance of the full plethora of nonnegative polynomials 
has been missed in the physics of 
nonclassical light in the last two decades~\cite{PhysRevLett.94.153601, PhysRevA.71.011802, PhysRevA.72.043808}.
Contrary to that, we show that the entire set of nonnegative polynomials allows us to reveal the nonclassicality of light optimally, even when they are hidden to the existing methods. \new{In this way, we discover a fundamentally complete analysis technique in the characterisation of nonclassical effects in \emph{finite} experimental data.} 
Moreover, addressing the problem at the abstract level allows us to rejoin the methodology for the investigation of the nonclassicality of light and spin systems. Specifically, classes of witnesses of nonclassicality for the latter can be mapped to those for the former and vice versa.
On the mathematical side, we introduce a hierarchy of semidefinite programs characterising the set of nonnegative bivariate polynomials from inside, which is in certain situations more convenient than the standard one for detecting nonclassicality of systems of many spins. 
We further introduce an outer approximation for the set of nonnegative bivariate polynomials, which allows for certifying that the nonclassicality is absent from the given data.

\begin{figure*}
    \centering
    \includegraphics[width=0.9\linewidth]{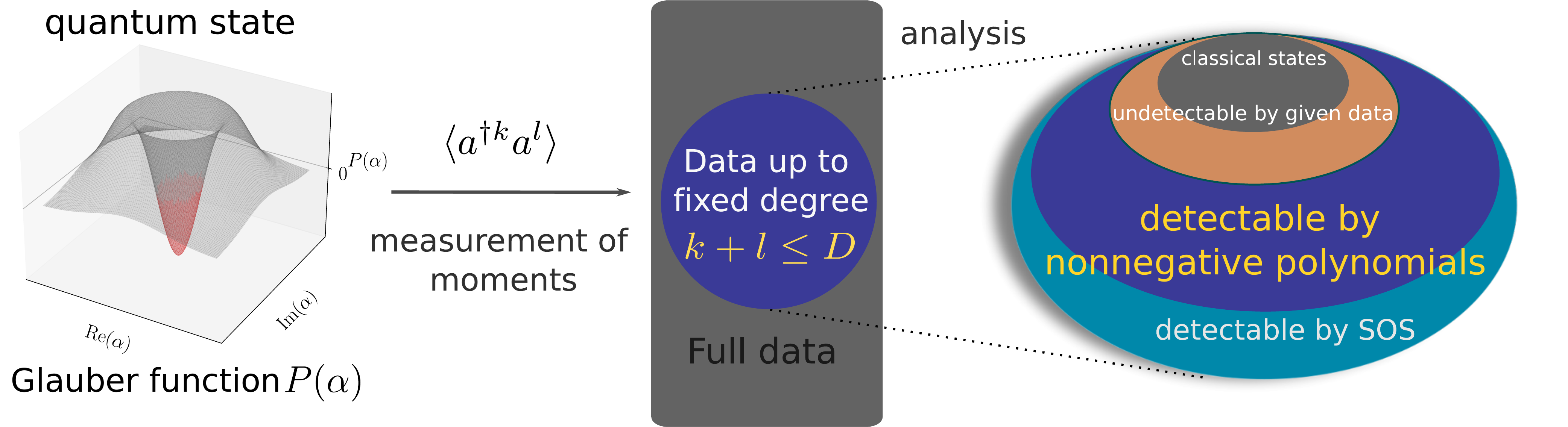}
    \caption{\new{Sketch of our approach to nonclassicality detection with finite data. Given a quantum state, the expectation values $\langle a^{\dagger k} a^{l} \rangle$ up to  degree $D$, i.e. $k+l \leq D$, are obtained experimentally. Analysis with nonclassicality witnesses by nonnegative polynomials of degree less than $D$ gives the exact characterisation of data which have nonclassical effects; the set of states detectable in this way is complementary to those that are fundamentally undetectable based on the given moment data. In contrast, the methods of sum of squares and moment matrix give an outer approximation to this set.}} 
    \label{fig:front}
\end{figure*}

\textit{Nonclassicality of light.---}
Classically, a light wave is the harmonic oscillation of the electromagnetic field in space and time. 
In quantum theory, such oscillation can be identified with the coherent states of a harmonic oscillator $\ket{\alpha}$, parameterised by a complex number $\alpha \in \mathbb{C}$, whose amplitude and phase correspond to those of the classical dynamics \cite{PhysRev.131.2766, Kam2023}.
One arrives at the general definition of classical light $\rho$ as probabilistic mixture of coherent states \cite{Mandel1995},
\begin{equation}
\label{eq:p_function}
    \rho = \int d^2 \alpha \, P(\alpha) \ket{\alpha} \bra{\alpha},
\end{equation}
where the Glauber-Sudarshan function $P(\alpha)$~\cite{PhysRevLett.10.277} is constrained to be a probability distribution on $\mathbb{C}$. 
Experiments demonstrating nonclassicality of a state~\cite{short_observation_1983, PhysRevA.78.021804, PhysRevLett.120.253602, Sperling2015, Andersen2016} can be formulated abstractly as estimating a witness of nonclassicality. 
Such a witness is an observable $W$ that has nonnegative expectation values for all coherent states, i.e., $\bra{\alpha} W \ket{\alpha} \geq 0$. Hence, $\mean{W}_\rho = \operatorname{tr} (\rho W) \ge 0$ holds for all classical states~\eqref{eq:p_function}, and 
a negative expectation value of the witness immediately implies that the state is nonclassical~\cite{PhysRevLett.94.153601}.

\textit{Nonnegative polynomials to detect nonclassicality of light.---}
Let $X$ and $P$ be the two quadrature operators subject to the commutation relation $[X,P]= i/2$ in units where $\hbar=1$. 
Any witness can be expanded in a power series of the quadratures, or equivalently in the creation and annihilation operators $a^{\dagger}= X - i P$ and $a=X + i P$, i.e., $W = \sum_{k,l} w_{kl} \, {a^{\dagger {k}}} a^l $~\cite{Weinberg1995-pi}.
Being a physical observable, $W$ is hermitian, implying that $w_{kl}=\bar{w}_{lk}$.
Notice that $p_{W}(\bar{\alpha},\alpha) := \bra{\alpha}W\ket{\alpha}$ is a real-valued function in $\alpha$ and its complex conjugate $\bar{\alpha}$. 
That $W$ is a nonclassicality witness is equivalent to $p_{W} (\bar{\alpha},\alpha)$ being nonnegative, anticipating the importance of nonnegative polynomials. 
However, if \emph{all} $\mean{W}_{\rho}$ can be computed from experimental data, one can restrict to a strict subset of nonnegative polynomials that are sum of squares of polynomials 
(hereafter, SOS)~\cite{PhysRevLett.94.153601,PhysRevA.71.011802, PhysRevA.72.043808}. 
Indeed, given the ability to compute $\mean{W}_{\rho}$, one considers
\begin{equation}
\label{eq:nonclassicality_optimisation_full}
          v = \underset{W}{\min} \{ \mean{W}_{\rho} : p_{W}(\bar{\alpha},\alpha) \geq 0 \}, 
\end{equation}
where $v < 0$ implies the nonclassicality of $\rho$. 
Additional linear constraints on $W$ can be added, allowing to quantify nonclassicality; see the supplemental material \cite[App.~A]{supplements_arxiv}. 
One then observes that any nonnegative function $p_{W} (\bar{\alpha},\alpha)$ can be written as $p_{W} (\bar{\alpha},\alpha) = \sqrt{p_{W} (\bar{\alpha},\alpha)}^2$, and $\sqrt{p_{W} (\bar{\alpha},\alpha)}$ can be approximated by a polynomial with sufficiently high degree~\cite{deBranges1959,PhysRevLett.94.153601}.
Accordingly, the optimisation~\eqref{eq:nonclassicality_optimisation_full} over all witnesses can be approximated by those that are SOS with sufficiently high degree~\cite{PhysRevLett.94.153601}.
SOS is an effective certificate for nonnegativity provided by a semidefinite program (SDP), a class of convex optimisation problems well-known in quantum physics~\cite{Skrzypczyk2023-li}; see also \cite[App.~B]{supplements_arxiv}.

This argument pertains to 
the standard methods for detecting nonclassicality by SOS ~\cite{PhysRevLett.94.153601} and by moment matrices developed in parallel~\cite{PhysRevA.72.043808,PhysRevA.71.011802, Richter2003, PhysRevLett.89.283601}. 
The latter amounts to measuring a matrix of observables constructed in a specific way, 
of which negative eigenvalues indicate nonclassicality~\cite[App.~C]{supplements_arxiv}.

However, any experimental design contains a finite number of measurement settings, and not all expectation values of witnesses $\mean{W}_\rho$ but only a subset thereof can be estimated with high accuracy~\cite{short_observation_1983, PhysRevA.78.021804, PhysRevLett.120.253602, Sperling2015, Andersen2016, PhysRevA.92.011801}.
\new{Indeed, moments $\mean{{a^{\dagger {k}}} a^l}$ are hard to obtain for high $k$ and $l$ due to the number of measurement settings that rises with the degree $k+l$~\cite{PhysRevA.72.043808}.
In this work, we show how to optimally exploit the data of a fixed polynomial degree to detect nonclassicality. This is achieved by accessing all nonnegative polynomials of the respective degree and thus going beyond the SOS approximation introduced in Ref~\cite{PhysRevLett.94.153601}. 
}   

Suppose that only data of moments $\mean{{a^{\dagger k} } a^l}_\rho$ up to a certain degree $k+l \le D$ are available.
Witnesses $W$ whose expectation values can be computed from the data are thus polynomials in $a^{\dagger}$ and $a$ and therefore $p_W (\bar{\alpha},\alpha)$ are \emph{polynomials}~ \cite[App.~A]{supplements_arxiv}. 
Taking into account this experimental constraint, the optimisation~\eqref{eq:nonclassicality_optimisation_full} becomes
\begin{equation}
\label{eq:nonclassicality_optimisation_problem}
     \begin{array}{lll}
          v_{D} = &\underset{W}{\min} &\mean{W}_{\rho}  \\
          & \textnormal{s.t} \; & p_W (\bar{\alpha},\alpha) \geq 0, \; \deg(p_{W}) \leq D.
     \end{array}
\end{equation}
Similarly to~\eqref{eq:nonclassicality_optimisation_full}, additional linear constraints can be added so that $v_D$ quantifies nonclassicality~\cite[App.~A]{supplements_arxiv}.
Notice that $v_D$ is an upper bound for $v$ which \emph{optimally} exploits the available data in the sense that any better upper bound requires more moments to be measured. \new{Physically, nonclassical states for which $v_D \geq 0$ are  not distinguishable from some corresponding classical states whose moments take the same values.} 
Unlike in the mentioned standard methods~\cite{PhysRevLett.94.153601, PhysRevA.71.011802}, the polynomial $p_W (\bar{\alpha},\alpha)$ in~\eqref{eq:nonclassicality_optimisation_problem} is constrained to be nonnegative, and not necessarily a SOS.

\textit{Reznick's hierarchy.---} 
Even if a given polynomial $p_W (\bar{\alpha},\alpha) > 0$ is not a SOS, it can be shown that $(1+ \bar{\alpha} \alpha)^b p_{W}(\bar{\alpha},\alpha)$ is a SOS for sufficiently large $b \in \mathbb{N}$~\cite{Reznick1995}.
Constraining $(1+ \bar{\alpha} \alpha)^b p_{W}(\bar{\alpha},\alpha)$ to be SOS, as mentioned above, can be formulated as a SDP in its parameters $w_{kl}$. This results in the first approach to problem~\eqref{eq:nonclassicality_optimisation_problem} in terms of a hierarchy of SDPs, \new{see the \hyperref[sec:end_matter]{{\color{black} End Matter}} and \cite[App.~B]{supplements_arxiv} for a detailed description.} 
It is now seen that the  approach developed by Ref.~\cite{PhysRevLett.94.153601, PhysRevA.72.043808,PhysRevA.71.011802, Richter2003, PhysRevLett.89.283601} corresponds to the zero-th level ($b=0$);
to fully exploit the same experimental data of moments of \textit{fixed} degree $D$, Reznick's hierarchy increases the exponent $b$. \new{Notice that the set of states that are undetectable for polynomial degree $D$ and Reznick order $b$ form a convex set and for high $D$ or $b$ the nonclassicality may be seen as hidden to the existing methods \cite{PhysRevLett.94.153601, PhysRevA.71.011802,PhysRevA.72.043808,Richter2003, PhysRevLett.89.283601}.} 

\begin{figure}[t]
    \includegraphics[width=0.36\textwidth]{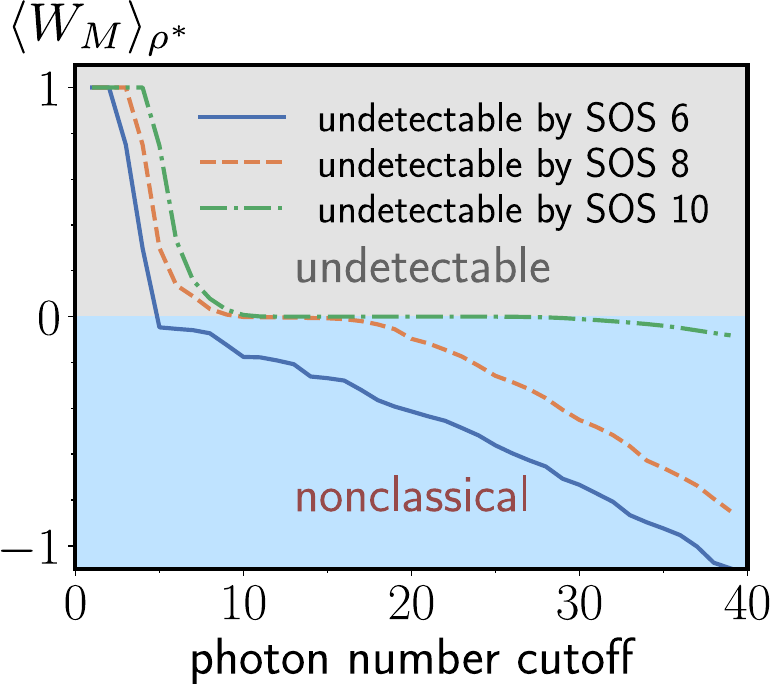}
    \caption{Nonclassicality indicated by the most negative expectation values \new{$\mean{W_M}_{\rho^{\ast}}$,  using the fixed Motzkin polynomial,} among all states that are undetectable by SOS up to degree $6$ (solid blue), $8$ (dotted orange), $10$ (dash-dot green). The $x$-axis indicates the photon number cutoff 
    of the state.} 
    \label{fig:motzkin}
\end{figure}

\textit{Examples.---} 
To demonstrate the importance of nonnegative polynomials which are not SOS in detecting nonclassicality, we exemplarily consider the famous Motzkin polynomial~\cite{motzkin1967arithmetic}. The corresponding witness $W_M$ is 
\begin{equation}
    W_{M} = \normord{X^{4}P^{2} + X^{2}P^{4} - 3 X^{2}P^{2} + \mathds{1}},
\end{equation}
where $\normord{O}$ for any observable $O$ denotes the normal reordering so that $a^\dagger$ is to the left of $a$. 
While $p_{W_{M}} (\bar{\alpha},\alpha)$ is not SOS, it is nonnegative as  $(1+ \bar{\alpha} \alpha) p_{W_{M}} (\bar{\alpha},\alpha)$ is a SOS, in accordance with Reznick's hierarchy.  
\new{We construct nonclassical states  having negative $\mean{W_M}_{\rho}$, but constrain their nonclassicality to be undetectable by SOS polynomials of degree $6, 8$ and $10$; see \cite[App.~D]{supplements_arxiv} for the details.
Figure \ref{fig:motzkin} illustrates the \new{minimal} values $\mean{W_M}_{\rho^\ast}$ obtained by a state $\rho^\ast$ among those, depending on the maximal number of photons allowed in its support.} 
The obtained states with only small photon number cutoff have $\mean{W_M}_\rho \ge 0$ and thus have also no nonclassicality detected by the Motzkin polynomial. 
However, deep negative values indicating strong nonclassicality can be observed for increasing photon number cutoff as seen in Fig.~\ref{fig:motzkin}.

\textit{Incorporating other experimental constraints---} 
The strength of the nonclassicality detection method as formulated in~\eqref{eq:nonclassicality_optimisation_problem} lies in its flexibility in incorporating different experimental constraints. 

Suppose that only moments of the photon number $\hat{n}=a^\dagger a$ up to a certain degree are measured.
Therefore, only witnesses $W$ that are polynomials in $a^\dagger a$ can be estimated. 
It then follows that  $p_W(\bar{\alpha},\alpha)$ is a univariate polynomial in $|\alpha|$, whose nonnegativity is equivalent to SOS \cite{Hilbert1888}. 
This allows for expressing the nonclassicality condition as a SDP, which includes the sub-poissonian statistics criterion $\mean{\hat{n}^2} - \mean{\hat{n}}^2 < \mean{ \hat{n}}$
as special case; see~\cite{PhysRevA.71.011802,PhysRevA.72.043808,PhysRevA.46.485, Richter2003, PhysRevLett.89.283601}. Notice that the mentioned results by Hilbert imply that this criterion exploits the available data optimally.
A related nonclassicality criterion based on photon counts~\cite{Innocenti_2022, Klyshko_1996} can be recovered in a similar way~\cite[App.~E]{supplements_arxiv}.  

As another example, we consider the case of moments of quadratures measured in several directions. 
Notice that this data is in general not sufficient to construct the minors of moment matrices, which requires also observables of the form $\mean{\normord{X^{k}P^{l}}}$~\cite{PhysRevA.71.011802}.
In particular, if moments of a single quadrature, say $X$, to a certain degree are measured, the corresponding polynomial $p_W (\bar{\alpha},\alpha)$ can be expressed in a single real variable $\Re(\alpha)$ and its positivity is equivalent to being a SOS. 
Thus, the optimal condition for nonclassicality can be expressed as a SDP, which includes the squeezing condition $\mean{X^2} - \mean{X}^2 < 1/4$~\cite{fox2006quantum} as special case. 
More interestingly, even if moments of two quadratures $X$ and $P$ are measured, we show that the optimal condition is decomposed into that for $X$ and $P$ separately, see \cite[App.~E]{supplements_arxiv}.

\textit{Nonclassicality of spin-$\frac12$ systems---}
The relationship between harmonic oscillators and systems of spins has been a long-standing inspiration in physics \cite{Jordan1935a,Holstein1940a,Schwinger1952a,PhysRevA.6.2211,Kam2023}.
Important for us is the similarity between the displacements on the plane of coherent states and collective rotations of a system of $m$ uncoupled spin-$\frac{1}{2}$ particles, which led to the definition of coherent states of the latter~\cite{Radcliffe1971,PhysRevA.6.2211,Kam2023}.  
Such spin coherent states are of the form $\ket{\psi^{\otimes m}}$, i.e., tensor powers of a pure qubit state. 
The classical states are then defined by~\cite{PhysRevA.78.042112}
\begin{equation}
    \rho = \sum_{\lambda} p_{\lambda} \ket{\psi_{\lambda}^{\otimes m}} \bra{\psi_{\lambda}^{\otimes m}},
    \label{eq:bosonic-sep}
\end{equation}
where $p_{\lambda}$ is a probability distribution.
The definition~\eqref{eq:bosonic-sep} turns out to coincide with that of the non-entangled states in the symmetric subspace~\cite{Eckert2002,Ghne2009, RevModPhys.81.865}. 

A spin nonclassicality witness $V$ is an observable such that 
\begin{equation}
\label{eq:bosonic_entanglement_witness}
   \bra{\psi^{\otimes m}} V \ket{\psi^{\otimes m}} \geq 0
\end{equation}
for all qubit vectors $\ket{\psi}$.
For a state $\rho$ in the spin system, $\min_{V} \mean{V}_\rho < 0$ implies that $\rho$ is nonclassical as defined in~\eqref{eq:bosonic-sep}. In practice, the constraint $\Tr(V)=m$ is also added for quantification of nonclassicality; see \cite[App.~F]{supplements_arxiv}. 
Expressing pure states $\ket{\psi}$ as points $(x,y,z)$ on the Bloch sphere, one obtains $p_V (x,y,z) = \bra{\psi^{\otimes m}} V \ket{\psi^{\otimes m}}$ as a polynomial over the $2$-sphere $\mathbb{S}^2$, with $x=\bra{\psi}\sigma_{x}\ket{\psi}$ for the Pauli matrix $\sigma_x$, and analogously for $y$ and $z$. 

Although the definition of spin system classicality~\eqref{eq:bosonic-sep} was motivated by that of light~\eqref{eq:p_function}~\cite{PhysRevA.78.042112}, the systems are different and methods for each are developed fairly independently~\cite{PhysRevA.78.042112, PhysRevA.86.042108,PhysRevA.100.062114,Wang2002,Tth2007,Qian2020}.
In fact, spin nonclassicality witnesses have been investigated by the method of nonnegative polynomials on the sphere~\cite{PhysRevA.69.022308, doherty2012convergence,deKlerk2020,Fang2020}.
Here, by formulating the characterisation of the nonclassicality of light using nonnegative polynomials, we can establish an exact mathematical mapping, which connects nonclassicality witnesses of spin systems to those of light.

\textit{Stereographic projection and the correspondence to nonclassicality of light---}
The stereographic projection maps a point $\beta \in \mathbb{C}$ to a point $(x,y,z) \in \mathbb{S}^2$ via \cite{Coxeter1989-mp} 
\begin{equation}
    (x,y,z) = \frac{1}{1 + \vert \beta \vert^2}(2\Re \beta,2\Im \beta,1-\vert \beta \vert^2).
\end{equation}
With this assignment one can associate $p_V(x,y,z)$ with a polynomial over the complex plane $\tilde{p}_{V}(\bar{\beta},{\beta})$, given by  $\tilde{p}_{V}(\bar{\beta},\beta)=(1+\vert \beta \vert^2)^{\deg(p_{V})} p_{V}(\bar{\beta},\beta)$. 
The polynomial $\tilde{p}_{V}(\bar{\beta},\beta)$ is nonnegative if and only if $p_{V} (x,y,z)$ is nonnegative on $\mathbb{S}^2$ establishing a quantitative relationship between the nonclassicality of the two different physical systems.

To illustrate, we consider the fidelity-based witness for spins~\cite{PhysRevLett.92.087902} given by $V_{\textnormal{GHZ}}:=\mathds{1}/2 - \ket{\textnormal{GHZ}}\bra{\textnormal{GHZ}}$ where $\ket{\textnormal{GHZ}} \propto \ket{0^{\otimes m}} + \ket{1^{\otimes m}}$ denotes the GHZ state~\cite{Greenberger1989}. 
Via the stereographic projection, the resulting bivariate polynomial $\tilde{p}_{V_{\textnormal{GHZ}}}(\bar{\beta},\beta)$ is translated to a witness $W_{\textnormal{GHZ}}$ for nonclassical light given by 
\begin{equation}
    W_{\textnormal{GHZ}} = - \frac{1}{2}(a^{m}+{a^{\dagger}}^{m}) + \frac{1}{2}\sum_{l=1}^{m-1}\binom{m}{l} {a^{\dagger}}^{l}a^{l}. 
\end{equation}
\new{When choosing $m=2$, $W_{\textnormal{GHZ}}$ is equal to the well-known squeezing witness $2P^2 - \frac{1}{2}\mathds{1}$ \cite{Andersen2016}. In general for fixed $m$, $W_{\textnormal{GHZ}}$ detects nonclassicality that stems from strong coherences between photon number states whose levels differ by $m$.}

As another example, it can be seen that the nonclassicality of Dicke-diagonal states is described by univariate polynomials and hence by SDPs in analogy to photon number witnesses for light, which was recognised rather recently~\cite{Yu2016,Quesada2017}.
 
In general, let $\mathcal{S}_{m}$ denote the set of nonclassicality witnesses for light obtained via the stereographic projection from a $m$-qubit system. 
We prove the following relationship 
\begin{equation}
    \mathcal{P}_{m} \subsetneq \mathcal{S}_{m} \subsetneq \mathcal{P}_{2m},
\end{equation}
where, $\mathcal{P}_{l}$ denotes the nonclassicality witnesses for light by polynomials of degree $l$; see \cite[App.~F]{supplements_arxiv}. \new{This implies that every nonclassicality witness of light can be obtained from a corresponding spin witness via the stereographic projection.}

\textit{An alternative for Reznick's hierarchy---} 
For experiments with measurements of high moments, the computation in Reznick's hierarchy can quickly become intractable.  
One such example is among the nonclassical states over $17$ spins constructed in~Ref.~\cite{Tura2018separabilityof}; \new{see the \hyperref[sec:end_matter]{{\color{black} End Matter}} for the description of the state}. 
 Reznick's hierarchy up to $b=9$ failed to detect its nonclassicality, and higher levels suffer from numerical instability. 
We address this problem by deriving an alternative hierarchy for the case of bivariate polynomials which combines two mathematical theorems on nonnegative polynomials, the P\'olya theorem~\cite{Polya_original}
and the Fejér-Riesz theorem~\cite{Fejér1916}. 
\begin{figure}[t!]
    \centering
    \includegraphics[width=0.4\textwidth]{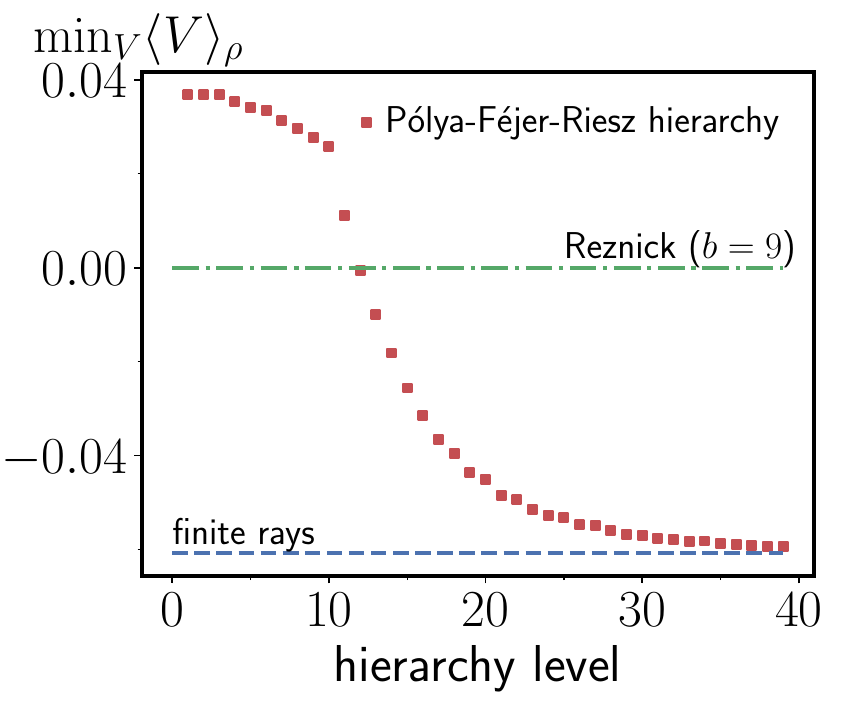}
    \caption{
    Detection of the $17$-qubit state in Eq.~\eqref{eq:seventeen_qubits_state} by the P\'olya-Fej\'er-Riesz hierarchy until level $40$ (red squares). Negative values approaching the lower bound obtained by the finite rays approximation (blue dashed line) indicate optimal detection. In contrast, the SDPs by Reznick's hierarchy  (until $b=9$) (green dash-dotted line) cannot detect the nonclassicality.}
    \label{fig:seventeen_qubits}
\end{figure}

To this end, we express the polynomial $\tilde{p}_V(\bar{\beta},\beta)$ in polar coordinates $\beta=r e^{i \theta}$ as 
$\tilde{p}_V = \sum_{s=0}^{2m} q_{s}(\theta) r^{s}$
where the coefficients $q_{s}(\theta)$ are trigonometric polynomials in $\theta$.  
P\'olya's theorem~\cite{Polya_original} implies that for large enough $b' \in \mathbb{N}$, the polynomial $(1+r)^{b'} \tilde{p}_{V} = \sum_{s=0}^{2m+b'} q_{s}^{(b')}(\theta) \, r^s$ has only nonnegative coefficients, $q_{s}^{(b')}(\theta) \geq 0$. 
The Fejér-Riesz theorem~\cite{Fejér1916} then allows to express the constraint $q_{s}^{(b')}(\theta) \geq 0$ as a SDP. 
With this, we obtain a hierarchy of SDPs for increasing $b'$ to characterise the set of nonclassicality witnesses,
referred to as the P\'olya-Fej\'er-Riesz hierarchy~\cite[App.~G]{supplements_arxiv}. 

While the P\'olya-Fej\'er-Riesz hierarchy can converge slower than Reznick's hierarchy, its complexity scales as $\mathcal{O}(m^{3}+m^{2} b')$ in contrast to $\mathcal{O}[(m+b)^{4}]$ for the latter; see \cite[App.~G]{supplements_arxiv}. One therefore can apply the hierarchy in cases when Reznick's hierarchy reaches the computational limit at hand. 
In particular, it is possible to detect the aforementioned state of $17$ qubits optimally, while Reznick's hierarchy fails to detect nonclassicality for $b\leq 9$ before it becomes numerically unstable, see Figure~\ref{fig:seventeen_qubits}.

\textit{Inner approximation by finite rays---}
Both Reznick's and the P\'olya-Fej\'er-Riesz hierarchy provide sequences of outer approximations for classical states. 
To practically estimate their optimality gap, we introduce also an inner approximation to the set of classical states based on an effective necessary condition for nonnegativity.  
We take a finite subset of rays specified by angles $\{\theta_{i}\}_{i=1}^{N}$ and demand the nonnegativity of the univariate polynomial $\tilde{p}_V(r,\theta_{i})$ for every fixed $\theta_i$ which can be expressed as SDP, see \cite[App.~H]{supplements_arxiv}.  
The approximation strength is clearly illustrated in Figure~\ref{fig:seventeen_qubits}, in particular demonstrating that the upper bounds of the nonclassicality based on the P\'olya-Fej\'er-Riesz hierarchy gradually converge to the obtained lower bound. 

\textit{Conclusion---}
Embedding the concept of nonclassicality in quantum mechanics in concrete experimental conditions, we have bridged the theory of nonnegative polynomials to the study of nonclassicality of light and spin systems at the same time. 
This uncovers an unexplored regime in the research of nonclassicality in quantum mechanics with questions that urge for further investigation. For example, what is the operational meaning of states that carry nonclassicality hidden to SOS \new{or higher orders in Reznick's hierarchy}, bearing its similarity to bound entangled states in entanglement theory?
Can multi-mode nonclassicality be treated with nonnegative polynomials?
The flexibility of the formulation of nonclassicality by nonnegative polynomials in incorporating experimental constraints is remarkable. 
In fact, extension of its applications to investigation of other problems where systems of many particles are involved can be expected.

\begin{acknowledgments}
The authors would like to thank 
Khazhgali Kozhasov,
Marco Túlio Quintino,
and
Geza T\'oth
for helpful discussions.
The University of Siegen is kindly acknowledged for enabling our computation through the \texttt{OMNI} cluster. 
This work was supported by 
the Deutsche Forschungsgemeinschaft (DFG, German Research Foundation, project numbers 447948357 and 440958198), 
the Sino-German Center for Research 
Promotion (Project M-0294), 
the ERC (Consolidator Grant 683107/TempoQ), 
and the German Ministry of Education 
and Research (Project QuKuK, BMBF 
Grant No. 16KIS1618K).
This project has received funding from the European Union's Horizon 2020 research and innovation programme under the Marie Sk\l odowska-Curie grant agreement No.~945422.
BO and TdW were partially funded by the German Federal Ministry for Economic Affairs and Climate Action (BMWK), project ProvideQ; TdW was moreover supported by the DFG grant WO 2206/1-1.

The source code for all numerical computations in this paper is available on a public repository~\cite{code}
\end{acknowledgments}


\vspace{1cm}
\section*{End Matter} \label{sec:end_matter}

\new{
\textit{Explicit description of Reznick's hierarchy for nonclassicality detection---} Here we provide an explicit formulation of the semidefinite programming hierarchy for nonclassicality detection of light using data of a fixed maximal degree $D \in 2\mathbb{N}$. First, let $\mathbb{Z}_{D}$ denote the set of integers $\{0,1,\dots,D\}$. Further, let $S \subset \mathbb{Z}_D \times \mathbb{Z}_D$ be a subset such that $k+l\leq D$ for all $(k,l) \in S$. We consider the case in which the data $S_{\rho} := \{\langle \normord{X^k P^l} \rangle_{\rho}: (k,l) \in S\}$ coming from an unknown state $\rho$ is given. Furthermore, let $\mathbf{m}_{d}$ denote the vector of bivariate monomials in the variables $x$ and $p$ of degree less or equal to $d \in \mathbb{N}$. To illustrate for $d=2$ the vector $\mathbf{m}_{2}$ is given by 
\begin{equation}
    \mathbf{m}_{2} = (1, x, p, x^2, xp, p^2)
\end{equation}
where $(\cdot)^T$ denotes the transpose.
To detect nonclassicality based on the data $S_{\rho}$, we optimise -- using Reznick's theorem -- over all nonnegative bivariate polynomials with exponents from $S$. Explicitly, with fixed exponent $b \in \mathbb{N}_{0}$, we obtain the semidefinite program
\begin{align*}
    &\textnormal{given} \; S_{\rho}, b \\
    &\min_{f_{kl}, G} \sum_{(k,l) \in S} f_{kl} \; \langle \normord{X^k P^l} \rangle_{\rho} \\
    &\textnormal{s.t.} \; (1+x^2+p^2)^b\sum_{(k,l) \in S} f_{kl} x^k p^l = \mathbf{m}_{\frac{D}{2}+b}^{T} \, G \, \mathbf{m}_{\frac{D}{2}+b} \\
    & \quad\;\;\; G \succcurlyeq 0
\end{align*}
where the inequality $\succcurlyeq$ refers to the cone of positive semidefinite matrices. If the outcome of this problem is negative, the state $\rho$ is detected as nonclassical and extracting the optimal coefficients $f_{kl}$ gives rise to a witness $W$ of nonclassicality via
\begin{equation}
    W = \sum_{(k,l)\in S} f_{kl} \normord{X^k P^l}
\end{equation}
In practice, to obtain a bounded result in case of detectability, one has to impose an additional affine normalisation constraint on $W$, for instance $\sum_{(k,l)\in S} f_{kl} = 1$.}

\new{\textit{Nonclassical state of $17$ qubits---} Here, we describe the state $\rho$ supported on the symmetric subspace of $17$ qubits that has been derived in Ref.~\cite{Tura2018separabilityof}. There, the authors proved that the state $\rho(a,b,K)$ of $m=2K+1$ qubits defined by 
\begin{equation}
 \label{eq:seventeen_qubits_state}
    \rho(a,b,K) = \frac{1}{2(4+a)^{K}}[\eta(a) + \gamma(b)]
\end{equation}
is nonclassical for all integers $K \geq 2$, $a\in(0,\infty)$ and $b\in \{\pm 1\}$. Here, the operators $\eta$ and $\gamma$ are explicitly given by 
\begin{align}
    \eta(a) &=  \sum_{l=0}^{m} \binom{m}{l} \lambda_{l}(a) \ket{m,l}\bra{m,l} \\
    \gamma(b) &= b(\ket{m,0}\bra{m,m} + \ket{m,m}\bra{0,m}).
\end{align}
The coefficients $\lambda$ are recursively defined via $\lambda_{l}(a) = f_{K-l}(a)$ with $f_{k+2}(a) = (2+a)f_{k+1}(a) - f_{k}(a)$ for all $k \in \mathbb{Z}$, $f_{0}(a)=1$ and $f_{1}(a)=1+a$. 
We especially considered the state $\rho(1,1,8)$ which is a symmetric state of $17$ qubits and computed the optimal witness violation by increasing the level in the P\'olya-Fej\'er-Riesz hierarchy, see Fig.~\ref{fig:seventeen_qubits}. 
}

\include{supplemental_arxiv}

\end{document}

%% file: supplemental_arxiv.tex
\appendix

\hspace{1.8cm}{\centering \bf Supplemental Material}
\section{Basic notations of quantum optics and nonclassicality of light}
\label{app:coherent_states}
Mathematically, a single mode of light is described by an excitation of a quantum harmonic oscillator in one dimension~\cite{fox2006quantum}. 
A basis of the Hilbert space 
of the system is given by the Fock states $\{\ket{n}\}_{n=0}^{\infty}$, where $n$ denotes the number of photons occupying the mode. 
The  creation operator $a^{\dagger}$ and the annihilation operator $a$  with the canonical commutation relation $[a,a^\dagger]=1$ act in this basis as 
\begin{equation}
    a \ket{n} = \sqrt{n} \ket{n-1}, \;\;\;\;\;\;\;\; a^{\dagger} \ket{n} = \sqrt{n+1} \ket{n+1}.
\end{equation}
The hermitian operators $X$ and $P$, known as the field quadrature observables, are defined in terms of $a$ and $a^{\dagger}$ as 
\begin{equation}
    X=\frac{a + a^{\dagger}}{2}, \;\;\;\;\;\;\;\; P=\frac{a - a^{\dagger}}{2i},
\end{equation}
where the units are chosen such that all physical parameters (mass and frequency) and the reduced Planck constant $\hbar$ are equal to $1$.  
A coherent state $\ket{\alpha}$ associated to the complex number $\alpha \in \mathbb{C}$ takes the form 
\begin{equation}
    \ket{\alpha} = e^{-\frac{\vert \alpha \vert^{2}}{2}} \sum_{n=0}^{\infty} \frac{\alpha^{n}}{\sqrt{n!}} \ket{n}.
    \label{eq:coherent-n}
\end{equation}
One can readily confirm that the state~\eqref{eq:coherent-n} is the normalised eigenstate of the annihilation operator $a$ with eigenvalue $\alpha$, 
\begin{equation}
    a \ket{\alpha} = \alpha \ket{\alpha}. 
\end{equation}

A general mixed state is described by a positive semidefinite trace-class operator $\rho$ with $\Tr(\rho) = 1$. Such a state $\rho$ is considered to be classical if it can be decomposed as a convex combination of coherent states [Eq. (1) in main text], i.e.,
\begin{equation}
    \rho = \int_{\mathbb{C}} d^2 \alpha \, P(\alpha) \ket{\alpha} \bra{\alpha}
    \label{eq:app-classical}
\end{equation}
where $d^2 \alpha \, P(\alpha)$ is some probability measure on $\mathbb{C}$. 

A witness of nonclassicality is defined as an operator $W$ such that $\bra{\alpha} W \ket{\alpha} \geq 0$ for all $\alpha \in \mathbb{C}$. 
It follows that $\mean{W}_\rho = \Tr (\rho W) \ge 0$ for all classical states~\eqref{eq:app-classical}.
We then consider [Eq. (2), main text]
\begin{equation}
\label{eq:nonclassicality_optimisation_full_app}
          v = \underset{W}{\min} \{ \mean{W}_{\rho} : p_{W}(\bar{\alpha},\alpha) \geq 0 \}. 
\end{equation}
Clearly, $v < 0$ implies that $\rho$ cannot be written as~\eqref{eq:app-classical} and is thus nonclassical by definition. 

In practice, it is also useful to consider the nonclassicality of a state $\rho$ \emph{quantitatively}. This can be obtained by slightly modifying the problem~\eqref{eq:opt_for_harm_osc}.
To this end, we consider
a convex combination $\rho_t$ of the considered state $\rho$ and some chosen classical state $\rho_{0}$, 
\begin{equation}
    \rho_{t} = t\rho + (1-t)\rho_{0}
\end{equation}
with $t \ge 0$. A typical choice for such a state $\rho_{0}$ is
\begin{equation}
    \rho_{0} = \int_{\mathbb{C}} d \alpha \bra{\alpha} \rho \ket{\alpha} \ket{\alpha} \bra{\alpha}.
    \label{eq:obvious-classical}
\end{equation}
Starting as $t=0$, $\rho_t$ is classical by construction. 
One can ask, how large $t$ can be until $\rho_t$ becomes nonclassical.
This corresponds to the solution of the optimisation problem 
\begin{equation}
\label{eq:visibility_optimisation}
    \begin{array}{lll}
       &s (\rho_0) = &\max \; t \\
        & &\textnormal{s.t.}  \; \rho_t \textnormal{ is classical}. 
    \end{array}
\end{equation}
This problem is connected to the minimisation~\eqref{eq:nonclassicality_optimisation_full_app} over all witnesses $W$ with the additional constraint that $\Tr(W\rho_{0}) = 1$ as follows
\begin{equation}
\label{eq:non_class_optimisation_with_cons}
    \begin{array}{lll}
      v (\rho_0) = &\min &  \Tr(W\rho) \\
        & \textnormal{s.t.} &  p_{W}(\bar{\alpha},\alpha) \geq 0 \\
       & & \Tr(W\rho_{0}) = 1.
    \end{array}
\end{equation}
Consider the value $\Tr(W \rho_t)$ and  notice that the optimal $W$ with respect to the problem~\eqref{eq:non_class_optimisation_with_cons} must satisfy $\Tr(W \rho_{s}) = 0$. This leads to the equality
\begin{equation}
    v(\rho_0) = 1 - \frac{1}{s (\rho_0)}
\end{equation} 
connecting the two optimisation problems \eqref{eq:visibility_optimisation} and \eqref{eq:non_class_optimisation_with_cons}. 

\section{Nonnegative polynomials and Reznick's hierarchy}
\label{app:sum_of_squares}
Attempts to characterise the cone of nonnegative polynomials and its relation to sums of squares have a long mathematical tradition reaching back to 19th century; see, e.g., \cite{Hilbert1888,Reznick:SurveyHilbert17th}.
In the 21st century, the problem regained massive interest due to its connection to polynomial optimisation problems.
The goal of unconstrained polynomial optimisation is the minimisation of a multivariate polynomial $f\in \mathbb{R}[x_{1},\dots,x_{n}]$. 
This computational problem is equivalent to the question of global nonnegativity of $f$ due to the following reformulation
\begin{equation}
\label{eq:unconstrained_poly_opt}
    \begin{array}{ll}
        \underset{\lambda \in \mathbb{R}}{\max} & \lambda  \\
         \text{s.t.} & f(\mathbf{x}) - \lambda \geq 0 \text{ for all } \mathbf{x} \in \mathbb{R}^{n}. 
    \end{array}
\end{equation}
However, the certification of nonnegativity of multivariate polynomials is NP-hard \cite{Laurent:Survey}, and solving \eqref{eq:unconstrained_poly_opt} is hard in practice. 
The common approach to tackle the problem of nonnegativity of polynomials is the usage of certificates of nonnegativity, i.e., a condition that implies nonnegativity, which is easier to check than nonnegativity itself, and that is satisfied for a broad range of nonnegative polynomials. 
The oldest and most prominent certificate are sum of squares decompositions.

A polynomial $f\in \mathbb{R}[x_{1},\dots,x_{n}]$ is a sum of squares, abbreviated SOS, if it can be decomposed as a finite sum
\begin{equation}
    f(\mathbf{x}) = \sum_{j=1}^{N} g_{j}(\mathbf{x})^{2},
\end{equation}
where $g_{j}\in \mathbb{R}[x_{1},\dots,x_{n}]$ are arbitrary polynomials. 
Obviously, every SOS is also nonnegative. 
However, the opposite implication is in general not true due to a seminal result by Hilbert in 1888 \cite{Hilbert1888}. 
Specifically, he proved that there exist exactly three classes of polynomials for which every nonnegative polynomial is a sum of squares:
\begin{enumerate}
    \item univariate polynomials,\\ e.g., \quad\quad $f(x) = x^6 + 2x^4 + x^2$,
    \item quadratic polynomials,\\ e.g., \quad\quad $f(x_{1},\dots,x_{n}) = \sum_{i=1}^{N} x_{i}^2$,
    \item bivariate quartic polynomials,\\ e.g., \quad\quad  $f(x,y) = x^4 + 6x^2 y + 9 y^2$.
\end{enumerate}
In all other classes, there exist polynomials which are nonnegative despite being not a SOS; see, e.g., the Motzkin polynomial, which was mentioned in the main text.

As briefly mentioned in the introduction, Hilbert's 1888 result lead to the 17th of his famous list problems in 1900, asking whether every nonnegative polynomial can be expressed as a sum of squares of rational functions. 
This is indeed the case due to a famous result by Artin from 1927 \cite{Artin:Hilberts17thProblem}.
Nowadays we know due to Reznick that we can even restrict to uniform denominators in those rational functions.

\begin{Theorem}[Reznick's Positivstellensatz; \cite{Reznick1995}]
\label{thm:Reznick-hierarchy}
Let $f\in \mathbb{R}[x_{1},\dots,x_{n}]$ be a polynomial such that $f(\mathbf{x}) > 0$ for all $\mathbf{x} \in \mathbb{R}^{n}\setminus \{0\}$. Then, there exists a number $b \in \mathbb{N}$ such that $f_b$ given by
\begin{equation}
    f_b(\mathbf{x}) := \left(\sum_{i} x_{i}^2 + 1\right)^{b} f(\mathbf{x}) 
\end{equation}
is a SOS.
\end{Theorem}
Theorem \ref{thm:Reznick-hierarchy} provides a sequence of SOS tests for a given polynomial and in the case that $f_b$ is a SOS for some $b$, a certificate for the nonnegativity of $f$ is constructed. 

Sums of squares are a very useful certificate, not only in theory but also in practice, as one can solve \eqref{eq:unconstrained_poly_opt} via semidefinite programming (SDPs), a well-known class of convex optimization problems; see e.g. \cite{Boyd:Vandenberghe}.
The key ingredients to see this are first that a polynomial is a sums of squares if and only if it admits a Gram matrix decomposition as follows.

\begin{Theorem}[Gram matrix method; \cite{Choi:Lam:Reznick, Powers:Woermann:GramMatrix}]
	Let $f \in \mathbb{R}[x_{1},\dots,x_{n}]$ be a polynomial of degree $2D$, and let $\mathbf{m}_D$ be the vector of all monomials in $n$ variables of degree at most $D$ (in lexicographic ordering). Then $f$ is SOS if and only if there exist a positive semidefinite matrix $G \geq 0$ such that
	\begin{equation}
		f \ = \ \mathbf{m}_D^T \cdot G \cdot \mathbf{m}_D. \label{Equation:GramMatrixMethod}
	\end{equation}
\end{Theorem}
Second, computing the maximal $\lambda \in \mathbb{R}$ such that $f - \lambda$ admits a Gram matrix representation as in \eqref{Equation:GramMatrixMethod} can be done by a SDP, which was observed by Parrilo \cite{Parrilo:Thesis} and around the same time by Lasserre \cite{Lasserre2000GlobalOW} using the language of moments. 
This approach gives an approximate solution maximizing $\lambda$ in \eqref{eq:unconstrained_poly_opt}, as one relaxes the condition of $f$ being nonnegative to $f$ being SOS.
Analogously, one can apply this method for computing bounds for the optimal $\lambda$ in \eqref{eq:unconstrained_poly_opt} using Reznick's hierarchy. 

The size of the vector $\mathbf{m}_{D}$ or equivalently the number of rows and columns of the corresponding Gram matrix $G$ is computed as $\binom{n + D}{D}$. Hence, in the case of bivariate polynomials ($n=2$), the number of scalar variables in the SDP's of Reznick's hierarchy scales as $\mathcal{O}((D+b)^{4})$.

\section{Witnesses of nonclassicality of light}
\label{app:harmonic}

\subsection{Witnesses of nonclassicality of light by nonnegative polynomials}

As a general ansatz for witness observables $W$ considered in the main text, we consider polynomials in $a^\dagger$ and $a$  of some degree $D \in \mathbb{N}$, i.e., hermitian operators $W$ of the form 
\begin{equation}
\label{eq:harm_witness_polynomial-a}
    W = \sum_{k+l \leq D} w_{k l}  {a^{\dagger k} a^l},
\end{equation}
with $w_{kl} \in \mathbb{R}$. 
Notice that the expansion~\eqref{eq:harm_witness_polynomial-a} is `normally ordered' so that $a^\dagger$ is on the left of $a$ in every term.
The expectation value of an observable of the form \eqref{eq:harm_witness_polynomial-a} in coherent states $\ket{\alpha}$ is hence described by a real-valued polynomial $p_{W}$ given by
\begin{equation}
\label{eq:harm_associated_polynomial-a}
    p_{W}(\Bar{\alpha}, \alpha) =  \sum_{k+l \leq D} w_{kl} \, \Bar{\alpha}^{k} \alpha^{l}.
\end{equation}
It is then clear that the observable $W$ of the form \eqref{eq:harm_witness_polynomial-a} gives nonnegative expectation values on all classical states if and only if the associated polynomial $p_W$ in Eq.~\eqref{eq:harm_associated_polynomial-a} is nonnegative for all  $\alpha \in \mathbb{C}$. 

The detection of nonclassicality with witness of the form~\eqref{eq:harm_witness_polynomial-a} can be stated as the optimisation problem [Eq. (3), main text]
\begin{equation}
\label{eq:opt_for_harm_osc}
     \begin{array}{lll}
          v_{D} = &\underset{W}{\min} &\mean{W}_{\rho}  \\
          & \textnormal{s.t} \; & p_{W}(\bar{\alpha},\alpha) \geq 0, \; \deg(p_{W}) \leq D,
     \end{array}
\end{equation}
Notice that $v_D < 0$ implies the nonclassicality of the quantum state. 
As we explain in the main text, we use Reznick's hierarchy to solve the optimisation problem  \eqref{eq:opt_for_harm_osc}  practically. The induced hierarchy of semidefinite programs gives optima $V_{D,b}$ (with $b \in \mathbb{N}$) converging to $v_{D}$, where
\begin{equation}
\label{eq:reznick_for_harm_osc}
     \begin{array}{ll}
          v_{D,b} = &\underset{W}{\min} \mean{W}_{\rho}  \\
          & \textnormal{s.t} \;\;  (1+\Bar{\alpha} \alpha )^{b} p_{W}(\Bar{\alpha},\alpha) \in \SOS \\  &  \deg(p_{W}) \leq D.
     \end{array}
\end{equation}

To connect with the method of moment matrices discussed below, it is also useful to have another representation of the witness using the quadrature operators $X$ and $P$,
\begin{equation}
\label{eq:harm_witness_polynomial}
    W = \sum_{k+l \leq D} \tilde{w}_{kl}  \normord{X^k P^l},
\end{equation}
where $\normord{ \cdot }$ is the normal ordering operation.
More precisely, the normal ordering on any monomials of $a^\dagger$ and $a$ simply permutes the positions of $a^\dagger$ and $a$ so that the former is on the left of the latter, e.g, $\normord{ \prod_{l=1}^{n} a^{r_{l}} {a^{\dagger}}^{s_{l}} } \; = \prod_{l=1}^{n} {a^{\dagger}}^{s_{l}} \prod_{l=1}^{n} a^{r_{l}}$. 
The action on a general operator $O$ is obtained by expanding $O$ in monomials of $a^\dagger$ and $a$ `freely', i.e.,  without imposing the relation $[a,a^\dagger]=1$ and applying the normal ordering to every monomial. 

For any operator $O$, it is straightforward to see that $\normord{XO} = 1/2 (a^\dagger \normord{O} +  \normord{O} a)$ and therefore $\bra{\alpha}\normord{ XO }\ket{\alpha} =  \Re (\alpha) \bra{\alpha}\normord{O}\ket{\alpha}$ as $a \ket{\alpha} = \alpha \ket{\alpha}$ and $\bra{\alpha} a^\dagger = \bar{\alpha} \bra{\alpha}$. Likewise, $\bra{\alpha} \normord{OP} \ket{\alpha} =  \Im (\alpha) \bra{\alpha}\normord{O}\ket{\alpha}$. As a result, we have
\begin{equation}
\label{eq:quadrature_coherent_expectation}
    \bra{\alpha}\normord{X^k P^l}\ket{\alpha} \; = \Re(\alpha)^{k} \Im(\alpha)^{l}
\end{equation}
and 
\begin{equation}
\label{eq:harm_associated_polynomial}
    p_{W}(\bar{\alpha},\alpha) =  \sum_{k+l \leq D} \tilde{w}_{kl} \, \Re(\alpha)^{k} \, \Im(\alpha)^{l}.
\end{equation}

As the last remark for this section, linear constraints on $W$ can be added in the same way as described in Appendix~\ref{app:coherent_states}, which allows for interpretation of the optimum $v_D$ in Eq.~\eqref{eq:opt_for_harm_osc} as quantification of nonclassicality.

\subsection{The method of moment matrices and SOS witnesses}
\label{app:ho_moment_matrices}
The method of moment matrices was developed by Lasserre in the mathematical context \cite{Lasserre2000GlobalOW}, and has since then been used widely. 
The relation between SOS and the methods of moments is given by duality.

In the context of nonclassicality detection, the first application of moment matrices is given in Refs. \cite{PhysRevA.46.485,PhysRevA.71.011802}. Per Refs.~\cite{PhysRevA.46.485,PhysRevA.71.011802}, the moment matrix $\mathcal{M}_{2D}(\rho)$ of a state $\rho$ of degree $2D$ built on the observables $X$ and $P$ is defined as 
\begin{equation}
    [\mathcal{M}_{2D}(\rho)]_{\textbf{m}, \textbf{n}} := \langle :X^{m_1 + n_{1}} P^{m_2 + n_2}: \rangle_{\rho},
\end{equation}
where $\textbf{m}=(m_1, m_2)$ and $\textbf{n}=(n_1, n_2)$ run over all monomial exponents of degree less or equal than $D$, i.e., $m_1 + m_2 \leq D$ and $n_1 + n_2 \leq D$. As an example, the moment matrix of order two is given by 
\begin{equation}
\label{eq:2_moment_matrix}
    \mathcal{M}_{2}(\rho) = \begin{pmatrix} 1 & \langle X \rangle_{\rho} & \langle P \rangle_{\rho} \\ \langle X \rangle_{\rho} & \langle :X^2: \rangle_{\rho} &  \langle :XP: \rangle_{\rho} \\ \langle P \rangle_{\rho} & \langle :XP: \rangle_{\rho} &  \langle :P^2: \rangle_{\rho} \end{pmatrix}.
\end{equation}
By the duality between SOS polynomials and positive semidefinite moment matrices, a nonclassical state $\rho$ is detectable by a SOS witness of degree $2D$ if and only if the moment matrix $\mathcal{M}_{2D}(\rho)$ has a negative eigenvalue. We provide here an elementary proof for the sake of self-containment. 

\begin{Proposition}
\label{prop:moment_matrix_sos}
    Let  $\mathcal{M}_{2D}(\rho)$ be the moment matrix of a state $\rho$ of the quantum harmonic oscillator  of degree $2D$. Then  $\mathcal{M}_{2D}(\rho)$ is positive-semidefinite if and only if $\Tr(W \rho) \geq 0$ for all observables $W$ with the associated polynomials $p_{W}(\bar{\alpha},\alpha)$ being SOS of at most degree $2D$.
\end{Proposition}
\textit{Proof: } We are going to show that the existence of a negative eigenvalue of $\mathcal{M}_{2D}(\rho)$ is equivalent to a SOS polynomial $p_W$ s.t. $\Tr(W \rho) < 0$. 

Assume that $\mathcal{M}_{2D}(\rho)$ has a negative eigenvalue. Let $\mathbf{v} = (v_{\mathbf{m}})_{\mathbf{m}}$ be the corresponding eigenvector. Then
\begin{equation}
\label{eq:sos_mom_equiv}
    \begin{array}{ll}
         0 &>\mathbf{v}^{T} \mathcal{M}_{2D}(\rho) \mathbf{v}  \\
         &= \sum_{\mathbf{m},\mathbf{n}} v_{\mathbf{m}} \,  \mathcal{M}_{2D}(\rho)_{\mathbf{m},\mathbf{n}} \, v_{\mathbf{n}}  \\
         &=   \Tr(\normord{\sum_{\mathbf{m},\mathbf{n}} v_{\mathbf{m}} v_{\mathbf{n}} X^{m_1 + n_1} P^{m_2 + n_2}} \rho) \\
         &=  \Tr[\normord{(\sum_{\mathbf{m}} v_{\mathbf{m}} X^{m_1} P^{m_2})^{2}} \rho].
    \end{array}
\end{equation}
Notice that $W = \normord{(\sum_{\mathbf{m}} \mathbf{v}_{\mathbf{m}} X^{m_1} P^{m_2})^{2}}$ is such that the polynomial $p_W (\bar{\alpha},\alpha)$ associated to $W$ via Eq.~\eqref{eq:harm_associated_polynomial} is the square polynomial given by
\begin{equation}
    p_W (\bar{\alpha},\alpha) = \left(\sum_{\mathbf{m}} \mathbf{v}_{\mathbf{m}} \Re(\alpha)^{m_1} \Im(\alpha)^{m_2}\right)^{2},
\end{equation}
which completes the first part. 

For the opposite direction, assume that there exists an observable $W$ such that $\Tr(\rho W)<0$ and $p_W(\bar{\alpha},\alpha)$ is a sum of squares, i.e.,
\begin{equation}
    p_W(\bar{\alpha},\alpha) = \sum_{l} p_{l}(\bar{\alpha},\alpha)^{2}.
\end{equation}
Necessarily, there must be at least one index $l$ such that the observable $W_{l}$ that corresponds to $p_{l}(\bar{\alpha},\alpha)^{2}$ satisfies $\Tr(W_{l} \rho)<0$. Assume now that $\mathbf{v}_{l}$ is the vector of coefficients of the polynomial $p_{l}(\bar{\alpha},\alpha)$. By the identity~\eqref{eq:sos_mom_equiv}, one can see that $\mathbf{v}_{l}^{T} \mathcal{M}_{2D}(\rho)\mathbf{v}_{l}$ is negative.  
$\Box$

\section{Nonclassicality beyond SOS with nonnegative polynomials}
\label{app:motzkin_polynomial}
Given a nonnegative bivariate polynomial $f(x,y)$ of degree $D \in \mathbb{N}$, one can formally construct a witness $W_f$ via~\eqref{eq:harm_associated_polynomial} by replacing monomials of the form $x^ky^l$ by $\normord{X^k P^l}$. For example, with the Motzkin's polynomial,
\begin{equation}
f_{M}(x,y) = x^4 y^2 + x^2 y^4 - 3 x^2 y^2 + 1,
\label{eq:motzkin-app}
\end{equation}
one obtains a witness [Eq. (4), main text]
\begin{equation}
   W_{M} = \normord{X^{4}P^{2} + X^{2}P^{4} - 3 X^{2}P^{2} + \mathds{1}}.
\end{equation}

If $f$ is a nonnegative, non SOS, bivariate polynomial of fixed degree $D$, one can explicitly construct nonclassical states which are detected by $W_f$ (the observable associated to $f$) but not by any SOS polynomial of some fixed degree $\widetilde{D}$ which can even be \emph{larger} than $D$. 
To do so, we simply consider the following optimisation problem:
\begin{equation}
\label{eq:non_sos_state_construction}
    \begin{array}{ll}
        \underset{\rho}{\min} & \Tr(W_f \rho) \\
         \text{s.t} & \rho \geq 0,\; \Tr(\rho) = 1 \\
         & \mathcal{M}_{\widetilde{D}}(\rho) \geq 0 .
    \end{array}
\end{equation}
If the outcome of the optimisation problem \eqref{eq:non_sos_state_construction} is negative, then the corresponding minimiser $\rho_{*}$ has the property of being detected by $W_f$ while not being detectable by any SOS polynomial of degree $\widetilde{D}$, by Proposition~\ref{prop:moment_matrix_sos}. 

In general, it is not convenient to minimise over all possible density matrices since the underlying Hilbert space is infinite-dimensional. 
Thus, one can restrict the problem \eqref{eq:non_sos_state_construction} to states which are cut off in the maximal photon numbers $n_{\max}$, i.e., restrict to states $\rho$ given by 
\begin{equation}
    \rho = \sum_{i,j=0}^{n_{\max}} \rho_{ij} \ket{i} \bra{j},
\end{equation}
where $\ket{i}$ and $\ket{j}$ are states of photon number $i$ and $j$ (Fock states), respectively.
\begin{figure*}
    \centering
\includegraphics[width=\textwidth]{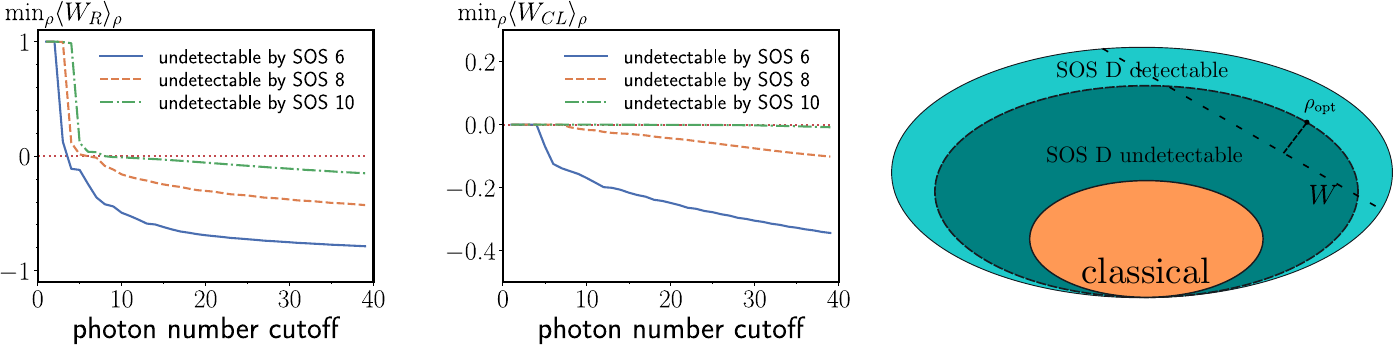}
    \caption{Plot of the expectation values of the observables $W_R$ and $W_{CL}$ corresponding to the Robinson and Choi-Lam polynomials minimized over all states that have nonnegative moment matrices of degree $6$ (solid blue), $8$ (dotted orange), $10$ (dashdot green). The $x$-axis corresponds to the maximal photon number in the support of the state.}
    \label{fig:robinson_choi_lam}
\end{figure*}

For the Motzkin polynomial~\eqref{eq:motzkin-app}, we evaluated the optimisation problem \eqref{eq:non_sos_state_construction} for different truncations of photon numbers specified by $n_{\max} \in \mathbb{N}$ and different positive semidefinite moment matrices indexed by $\widetilde{D}$ (see Fig.~2, main text). At $n_{\max}=10$ and $\widetilde{D}=6$, as an example, the obtained state reads as 
\begin{equation}
    \rho = p \ket{\psi_1}\bra{\psi_1} + (1-p) \ket{\psi_2}\bra{\psi_2},
\end{equation}
with 
\begin{align}
    \ket{\psi_1} &= a \ket{1} + b \ket{5} + c\ket{9} \\
    \ket{\psi_2} &= d \ket{2} + e \ket{6} + f\ket{10}.
\end{align}
The coefficients take the values $p = 0.06619, a=0.69896, b=-0.68135, c=-\sqrt{1-a^2 - b^2}, d=-0.94138, e=0.3124$ and $f=\sqrt{1-d^2 - e^2}$.

Similar observations as for the case of the Motzkin polynomial can be also made for different nonnegative polynomials which admit no SOS decomposition 
like the Robinson polynomial $f_R$ \cite{robinson1973some} or the Choi-Lam polynomial $f_{CL}$ \cite{choi1977old} 
\begin{align}
    f_{R}(x,y) &= x^6 - x^4 y^2 - x^2 y^4 + y^6 - x^4 + 3 x^2 y^2 \nonumber \\
    & \;\;\; - y^4 - x^2 - y^2 + 1,\\
    f_{CL}(x,y) &= x^4 y^2 - 3x^2 y^2 + y^4 + x^2. 
\end{align}
The observables $W_{R}$ and $W_{CL}$ associated to the polynomials $f_R$ and $f_{CL}$ may detect states which are undetectable by SOS polynomials of degree $6$, $8$ or even $10$, see Figure \ref{fig:robinson_choi_lam}.  For specification, all numerical computations in this work have been performed with a computer with a CPU Intel(R) Core(TM) i5-3570 processor (4 cores) running at 3.40GHz using 16 GB of RAM. 

Apart from these interesting properties regarding the detection of nonclassicality, nonnegative polynomials that are not SOS have been extensively studied in the mathematical literature.
In fact, both the Motzkin polynomial and the Choi-Lam polynomial are a special example of a broader class of polynomials first introduced as agiforms \cite{Reznick:AGI}, and then generalised as (sums of nonnegative) circuit polynomials (SONC) \cite{Iliman:deWolff:Circuits}. 
This class is (under mild technical adjustments) also referred to as SAGE~\cite{Chandrasekaran:Shah:SAGE}, and was independently (without another recurring terminology) discussed in the context of chemical reaction networks~\cite{Craciun:Pantea:Koeppl}. 
Circuit polynomials can be detected effectively via geometric programs \cite{Ghasemi:Marshall, Iliman:deWolff:GP,Dressler:Iliman:deWolff:FirstConstrained} or, more generally, relative entropy programs \cite{Chandrasekaran:Shah:SAGE,Chandrasekaran:Murray:Wiermann,Chandrasekaran:Murray:Wiermann2}, as well as (using a different approach) via second order cone programs \cite{Averkov,Wang:Magron} \textbf (which are all convex programs).
Moreover, they admit a Schm\"udgen-type Positivstellensatz \cite{Dressler:Iliman:deWolff:Positivstellensatz} (but not Putinar-type \cite{Dressler:Kurpisz:deWolff:SONCHypercube}), and they form a full-dimensional convex cone inside the cone of nonnegative polynomials \cite{Dressler:Iliman:deWolff:Positivstellensatz}, which has interesting duality aspects \cite{Dressler:Naumann:Theobald,Katthaen:Naumann:Theobald,Dressler:Heuer:Naumann:deWolff:DualSONCLP,Papp}, and a distinctive (algebraic) boundary structure \cite{Forsgaard:deWolff:BoundarySONCCone}.

\section{Incorporating other experimental constraints}
\label{app:alternative_schemes}
This appendix summarises how some particular well known experimental constraints for the detection of nonclassicality can be embedded into the framework of nonnegative polynomials. 

\subsubsection{Photon statistics and quadratures}
As the first example, the polynomial approach [Eq. (3) main text] assumes that all
moments $\langle{a^{\dagger {k}}} a^l \rangle$ up to some degree $D$ are experimentally accessible. As we discuss in the main text, additional constraints from experimental setup may require that the polynomials are further restricted to a certain subset of operators $a^{\dagger {k}} a^l$ indexed by tuples $(k,l)$ or equivalently a certain subset of quadrature observables $\normord{ X^{k}P^{l} }$. 

If only observables of the form $a^{\dagger {l}} a^l$  are taken into account so that $W=\sum_{l=0}^{D} w_{ll} \, {a^{\dagger {l}}} a^l$, which physically correspond to the analysis of photon number statistics, then the corresponding polynomial $p_W$ is given by 
\begin{equation}
\label{eq:fock_diagonal_witness_polynomial}
    p_W(\alpha, \Bar{\alpha}) = \sum_{l=0}^{D} w_{ll} \, \vert \alpha \vert^{2l}.
\end{equation}
Thus, checking if a polynomial in the photon number operator is a nonclassicality witness amounts to certifying the nonnegativity of a univariate polynomial in $\vert \alpha \vert$. If the polynomial is univariate, this is equivalent to testing whether it is a SOS due to Hilbert's 1888 result \cite{Hilbert1888}.

Similarly, in the cases where $W$ is a polynomial in the quadrature $X$ in the sense $W = \sum_{l=0}^{D} w_{l} \, \normord{ X^{l} }$, one obtains univariate polynomials of the form 
\begin{equation}
    p_W(\alpha, \Bar{\alpha}) = \sum_{l=0}^{D} w_{l} \ \Re(\alpha)^{l}
\end{equation}
and analogously for $P$ by replacing the real parts with imaginary parts. This induces a similar reducibility to SOS as in the case of measuring only the photon number. 

Another example we considered in the main text assumed that moments of the form $\langle \normord{X^l}\rangle$ and $\langle \normord{P^l}\rangle$ for $l\leq D$ are accessible for both quadrature observables. Interestingly, one can show that the joint data of $\langle \normord{X^l}\rangle$ and $\langle \normord{P^l}\rangle$ cannot detect any types of nonclassicality that can not be detected by the data  $\langle \normord{X^l}\rangle$ or $\langle \normord{P^l}\rangle$. This fact can be formally expressed by the following proposition.
\begin{Proposition}
    Let $D \in \mathbb{N}$, and $Q(\mathbf{q}) = \sum_{l=0}^{D} q_{l} \, \normord{ X^{l} }$, $R(\mathbf{r}) = \sum_{l=0}^{D} r_{l} \, \normord{ P^{l} }$ be families of observables parametrised by vectors $\mathbf{q}, \mathbf{r} \in \mathbb{R}^{D+1}$ and let $\rho$ be a state. Then, the optimisation problems
    \begin{equation}
    \label{eq:x_p_opt_1}
        \begin{array}{lll}
           a &=  \underset{\mathbf{q},\mathbf{r}}{\min} & \Tr(\rho W)  \\
              & \;\;\;\;\; \textnormal{s.t} &  W = Q(\mathbf{q}) + R(\mathbf{r}) \\
              && p_{W}(\bar{\alpha},\alpha) \geq 0 \\
        \end{array}
    \end{equation}
    and 
    \begin{equation}
     \label{eq:x_p_opt_2}
        \begin{array}{lll}
           b &=  \underset{\mathbf{q},\mathbf{r}}{\min} & \Tr(\rho W)  \\
              & \;\;\;\;\; \textnormal{s.t} &  W = \widetilde{Q}(\mathbf{q}) + \widetilde{R}(\mathbf{r}) \\
              && p_{\widetilde{Q}}[\Re(\alpha)] \geq 0 \\
              && p_{\widetilde{R}}[\Im(\alpha)] \geq 0
        \end{array}
    \end{equation} 
    result in the same optimum, $a=b$. Consequently, if the state $\rho$ is detectable by a witness $W = Q(\mathbf{q}) + R(\mathbf{r})$ then it is also detectable by a witness of the form $Q(\mathbf{q})$ or $R(\mathbf{r})$.
\end{Proposition}
\textit{Proof:} First it is easy to see that $b\geq a$ as the conditions $p_{Q}(\bar{\alpha},\alpha) \geq 0$ and $p_{R}(\bar{\alpha},\alpha) \geq 0$ are sufficient for $p_{W}(\bar{\alpha},\alpha) \geq 0$.

It is now to be seen that also $a \geq b$ holds. To this end, let $W_{*} = Q_{*} + R_{*}$ be the observable that minimises the optimisation problem \eqref{eq:x_p_opt_1}. The condition $p_{W_{*}}(\bar{\alpha},\alpha) \geq 0$ implies that $p_{Q_{*}} (\bar{\alpha},\alpha) - c \geq 0$ and $p_{R_{*}} (\bar{\alpha},\alpha) + c \geq 0$ are nonnegative polynomials for $c = \min p_{Q_{*}} (\bar{\alpha},\alpha)$. But since $W_{*} = Q_{*} - c \mathds{1} + R_{*} + c\mathds{1}$, we see that $W_{*}$ is also feasible in the optimisation problem \eqref{eq:x_p_opt_2} which shows that $a \geq b$. $\Box$

\subsubsection{Witnesses by photon number probabilities}

Besides the measurement quadrature observables to detect nonclassicality, criteria in terms of photon number probabilities have been derived in the past \cite{Innocenti_2022, Klyshko_1996}. Here, one assumes that photon number probabilities up to a maximal photon number $n_{\max} \in \mathbb{N}$, given by the expectation values $\{\mean{ \ket{n}\bra{n}} \}_{n=0}^{n_{\max}}$, are known. The question is whether these data are compatible with any classical state; if that is not the case, the prepared state has to be nonclassical. Consequently, given a state $\rho$, the task is to find an operator $W = \sum_{n=0}^{n_{\max}} w_{n} \ket{n}\bra{n}$ such that 
\begin{align}
    \bra{\alpha} W \ket{\alpha} &= \sum_{n=0}^{n_{\max}} w_{n} \, {\vert\braket{\alpha|n}\vert}^{2}\\
    &= e^{-\vert \alpha \vert^{2}} \sum_{n=0}^{n_{\max}} w_{n} \frac{\vert \alpha \vert^{2n}}{n!} \geq 0 \label{eq:exp_poly_non_neg}
\end{align}
for all $\alpha \in \mathbb{C}$. Of course, the nonnegativity constraint \eqref{eq:exp_poly_non_neg} is independent of the exponential factor $e^{-\vert \alpha \vert^{2}}$ so that the optimisation over witnesses involving only photon probability can be formulated as \begin{equation}
\label{eq:photon_number_prob_optmisation}
    \begin{array}{ll}
        \min_{w} & \sum_{n=0}^{n_{\max}} w_{n}  \mean{ \ket{n}\bra{n}}_{\rho}\\
         \textnormal{s.t} & p_{w}(x) = \sum_{n=0}^{n_{\max}} \frac{w_{n}}{n!} x^{2n} \geq 0 \textnormal{ for all } x \in \mathbb{R}.
    \end{array}
\end{equation} 
The problem \eqref{eq:photon_number_prob_optmisation} is again an optimisation over the set of nonnegative univariate polynomials that is closely related with the set of witnesses in terms of photon number moments described in Eq.~\eqref{eq:fock_diagonal_witness_polynomial}.

\section{Relationship between nonclassicality of light and that of spin systems}
\label{app:spin}
The intention behind this appendix is to provide some details on the relation between the nonclassicality witnesses of light and of spin systems. We saw that the witness condition of an observable $V$ for spins corresponds to the nonnegativity condition of a polynomial $p_V (x,y,z)$ in three variables constrained to the sphere $\mathbb{S}^2$. 
As we mentioned in the main text, in constrast to research on the nonclassicality of light, polynomial optimisation on the sphere has been considered in entanglement theory of spin systems ~\cite{PhysRevA.69.022308, doherty2012convergence,deKlerk2020,Fang2020}. 

Via the stereographic projection, one can construct an associated bivariate polynomial $\tilde{p}_{V}(\bar{\beta},{\beta})$ that has the property of being globally nonnegative if and only if $p_V$ is nonnegative on $\mathbb{S}^2$ as described in the main text. By the correspondence between nonnegative bivariate polynomials and nonclassicality witnesses of light, one can also directly construct a corresponding witness. 

Consequently, the construction of a nonclassicality witness of a symmetric $m$ qubit state $\rho$ is obtained by the optimisation problem
\begin{equation}
\label{eq:spin_classicality_optimisation_problem}
     \begin{array}{ll}
          l = &\underset{V}{\min} \mean{V}_{\rho}  \\
          & \textnormal{s.t} \; \tilde{p}_{V}(\beta,\bar{\beta}) \geq 0 , \Tr(V) = m
     \end{array}
\end{equation}
providing a strong similarity between the nonclassicality problem of light and of spin systems in the sense that both demand for a characterisation of nonnegative bivariate polynomials. 

For a deeper understanding of the similarity of both problems, it is useful to understand the structure of the polynomials $\tilde{p}_{V}(\beta,\bar{\beta})$. A convenient choice to represent these polynomials makes use of the Dicke-basis expansion $V = \sum_{ij=0}^{m} V_{ij} \ket{m,i}\bra{m,j}$ of the observable $V$. Given that a qubit vector $\ket{\psi}$ is parametrised as $\ket{\psi} = 1/\sqrt{1+\vert \beta \vert^{2}}(\ket{0} + \beta\ket{1})$ one can explicitly expand $\ket{\psi^{\otimes m}} = (1/\sqrt{1+\vert \beta \vert^{2}}^{m}) \sum_{l=0}^{m} \sqrt{\binom{m}{l}} \beta^{l} \ket{m,l}$.  

As a result, we see that the bivariate polynomial $\tilde{p}_{V}(\bar{\beta},{\beta})$ obtained by stereographic projection from a spin observable $V$ reads
\begin{equation}
\label{eq:stereographic_polynomial}
    \tilde{p}_{V}(\bar{\beta},{\beta}) = \sum_{ij=0}^{m} V_{ij} \sqrt{\binom{m}{i} \binom{m}{j}} \Bar{\beta}^{i} \beta^{j}.
\end{equation}
This allows to perform the stereographic projection on the level of polynomials in a very simple manner, once the Dicke basis expansion is known, as illustrated by the following example. 
\begin{Proposition}
    The nonclassicality witness of light obtained from the $m$ qubit GHZ-fidelity-witness $V_{\textnormal{GHZ}}:=\mathds{1}/2 - \ket{\textnormal{GHZ}}\bra{\textnormal{GHZ}}$ is 
    \begin{equation}
    \label{eq:ghz_light_witness}
    W_{\textnormal{GHZ}_{m}} = - \frac{1}{2}(a^{m}+{a^{\dagger}}^{m}) + \frac{1}{2}\sum_{l=1}^{m-1}\binom{m}{l} {a^{\dagger}}^{l}a^{l}. 
\end{equation}
\end{Proposition}
\textit{Proof:} The coefficients $\{V_{ij}\}_{ij=0}^{m}$ of $V_{\textnormal{GHZ}}$ in the Dicke basis are $V_{ii} = 1/2$ for $i=1,\dots,m-1$ and $V_{0m} = V_{m0} = -1/2$ and all other coefficients are equal to zero. Inserting these coefficients into Eq.~\eqref{eq:stereographic_polynomial} leads to the polynomial $\tilde{p}_{V}(\bar{\beta},{\beta})$ given by 
\begin{equation}
    \tilde{p}_{V}(\bar{\beta},{\beta}) = - \frac{1}{2}(\beta^{m}+\bar{\beta}^{m}) + \frac{1}{2}\sum_{l=1}^{m-1}\binom{m}{l} \bar{\beta}^{l} \beta^{l}
\end{equation}
which directly gives rise to the witness \eqref{eq:ghz_light_witness} by the substitution $\beta \mapsto a$ and $\bar{\beta} \mapsto a^\dagger$. $\Box$ 

Given the relationship, the following natural question arises: How is the set of polynomials in  Eq.~\eqref{eq:stereographic_polynomial} related to the space of all polynomials of a certain degree. 
In general, let $\mathcal{S}_{m}$ denote the set of bivariate polynomials obtained via the stereographic projection from operators of the $m$-qubit system and let $\mathcal{P}_{m}$ denote the set of all bivariate polynomials of degree $m$.
\begin{Proposition}
The following inclusions of sets of polynomials hold:
    \begin{equation}
    \mathcal{P}_{m} \subsetneq \mathcal{S}_{m} \subsetneq \mathcal{P}_{2m}.
\end{equation}
Consequently, nonclassicality witnesses of the harmonic oscillator, that are obtained from spin witnesses for $m$ spin-1/2's via stereographic projection, are nested between the sets of all nonclassicality witnesses with degree $m$ and $2m$. 
\end{Proposition}
\textit{Proof:} The sets $\mathcal{P}_{m}$ and $\mathcal{S}_{m}$ can represented as follows:
\begin{align}
    p(\Bar{\alpha}, \alpha) \in \mathcal{P}_{m} &\Leftrightarrow p(\Bar{\alpha}, \alpha) = \sum_{i+j\leq m} p_{ij} \, \Bar{\alpha}^{i} \alpha^{j}, \\
    p(\Bar{\alpha}, \alpha) \in \mathcal{S}_{m} &\Leftrightarrow p(\Bar{\alpha}, \alpha) = \sum_{i,j=0}^{m} p_{ij} \, \Bar{\alpha}^{i} \alpha^{j}.
\end{align}
From that, the inclusion $\mathcal{P}_{m} \subsetneq \mathcal{S}_{m}$ can be immediately seen noting that $i+j\leq m$ implies that $i\leq m$ and $j \leq m$. The inclusion is proper which can be seen from the example $p(\Bar{\alpha}, \alpha) = \vert \alpha \vert^{2m}$ that is a member of $\mathcal{S}_{m}$ but not of $\mathcal{P}_{m}$. 

The inclusion $\mathcal{S}_{m} \subsetneq \mathcal{P}_{2m}$ can be seen by noting that $i\leq m$ and $j \leq m$ together imply that $i+j \leq 2m$. The inclusion is also proper and that is demonstrated by the example $p(\Bar{\alpha}, \alpha) = \alpha^{2m} + \Bar{\alpha}^{2m}$. $\Box$



\section{A hierarchy derived from theorems by P\'olya and Fejér-Riesz}
\label{app:hybrid_hierarchy}
The starting point of the hierarchy is the polar decomposition of bivariate polynomials. 
A bivariate polynomial $f(x,y)$ of degree $2m$ has the form
\begin{equation}
    f(x,y) = \sum_{k+j \leq 2m} f_{kj} x^k y^j.
\end{equation}
If we interpret $x$ and $y$ as real and imaginary parts of a complex number $z$, in the sense that $x=(z+\Bar{z})/2$ and $y=(z-\Bar{z})/2i$, then we can write $f$ as a Hermitian polynomial $f(z,\Bar{z})$ given by 
\begin{equation}
    f(z,\Bar{z}) = \sum_{k+j \leq 2m} \widetilde{f}_{kj} \Bar{z}^{k} z^j
\end{equation}
with $\widetilde{f}_{kj} = \overline{\widetilde{f}_{jk}}$. In polar coordinates $z=re^{i\theta}$ with $r\in \mathbb{R}_{+}$ and $\theta \in [0,2\pi]$, this polynomial can be represented as 
\begin{align}
    f(r,\theta) &= \sum_{k+j \leq 2m} \widetilde{f}_{kj} e^{i(j-k)\theta} r^{k+j} \\
    &= \sum_{s=0}^{2m} \sum_{k+j=s} \widetilde{f}_{kj} e^{i(j-k)\theta} r^{s} \\
    &= \sum_{s=0}^{2m} q_{s}(\theta) r^{s} \label{eq:hybrid_representation}
\end{align}
where $q_{s}(\theta) := \sum_{k+j=s} \widetilde{f}_{kj} e^{i(j-k)\theta}$. 

The representation \eqref{eq:hybrid_representation} of the polynomial $f$ allows to construct a convergent hierarchy of semidefinite programs to characterise the set of positive bivariate polynomials. One of the main ingredients used for the construction of this hierarchy is P\'olya's theorem that characterises homogeneous polynomials that are positive for nonnegative variables.

\begin{Theorem}[P\'olya's theorem; \cite{Polya_original}]
\label{thm:polya}
    Let $f\in \mathbb{R}[x_{1},\dots,x_{n}]$ be a  polynomial that is positive on $\mathbb{R}_{+}^{n} \setminus \{0\}$. Then, there exists a number $b \in \mathbb{N}$ such that 
    \begin{equation}
    \label{eq:polya_polynomial}
        f_{b} = \left(\sum_{i=1}^{n} x_i + 1\right)^{b}f
    \end{equation}
    has only nonnegative coefficients.
\end{Theorem}
The other main building block is the characterisation of nonnegativity of real univariate trigonometric polynomials. A real univariate trigonometric polynomial of degree $t$ is given by a function $p:[0,2\pi] \rightarrow \mathbb{R}$ that can be represented as 
\begin{equation}
    q(\theta) = \sum_{l=-t}^{t} c_{l} e^{il\theta}
\end{equation}
where the coefficients $c_{l}\in \mathbb{C}$ satisfy $c_{l} = \overline{c_{-l}}$. The characterisation of nonnegative trigonometric polynomials is a widely studied subject and important in many areas of applied and general mathematics, see \cite{Dumitrescu2017} for a detailed introduction. The set of nonnegative univariate trigonometric polynomials admits a simple characterisation by the famous Fejér-Riesz theorem \cite{Fejér1916}. The Fejér-Riesz theorem states that a real univariate trigonometric polynomial $p$ is nonnegative if and only if it can be decomposed as $p=\vert g(\theta)\vert^2$ where $g(\theta)$ is function of the form
\begin{equation}
    g(\theta) = \sum_{l=0}^{t} g_{l} \, e^{il\theta},
\end{equation}
where $g_{l} \in \mathbb{C}$.
This means that every such polynomial is a hermitian square which is strongly reminiscent of the SOS characterisation of univariate (algebraic) polynomials discussed in Appendix~\ref{app:sum_of_squares}. 

As a direct consequence, the question whether a given real univariate trigonometric polynomial is nonnegative can be answered with a semidefinite program as stated by the following theorem.
\begin{Theorem}[Ref. \cite{Dumitrescu2017}]
\label{thm:pos_trig_polynomial}
    A real trigonometric polynomial $q(\theta) = \sum_{l=-t}^{t} c_{l} e^{il\theta}$ of degree $t \in \mathbb{N}$ is nonnegative for all $\theta \in [0,2\pi]$ if and only if there exists a complex positive semidefinite matrix $Q$ of size $(t+1) \times (t+1)$ such that 
    \begin{equation}
        c_{l} = \Tr(T_{l}Q)
    \end{equation}
    for all $l=-t,\dots,t$ where $T_{l}$ is the Toeplitz matrix defined as $(T_{l})_{mn}=1$ if $n-m=l$ and $(T_{l})_{mn}=0$ else. 
    In the case in which all $c_{l}'s$ are real, the matrix $Q$ can also be chosen to be real.  
\end{Theorem}

We are now ready to derive our hierarchy that is essentially a combination of the results given by Theorem~\ref{thm:polya} and Theorem~\ref{thm:pos_trig_polynomial}. 
The main idea comes from the observation that a bivariate polynomial $f$ can be thought of as a univariate polynomial constrained to the positive half-axis whose coefficients are real univariate trigonometric polynomials. This observation can be made with regard to the representation~\eqref{eq:hybrid_representation}. 

 \begin{figure}
     \centering
     \includegraphics[width=0.4\textwidth]{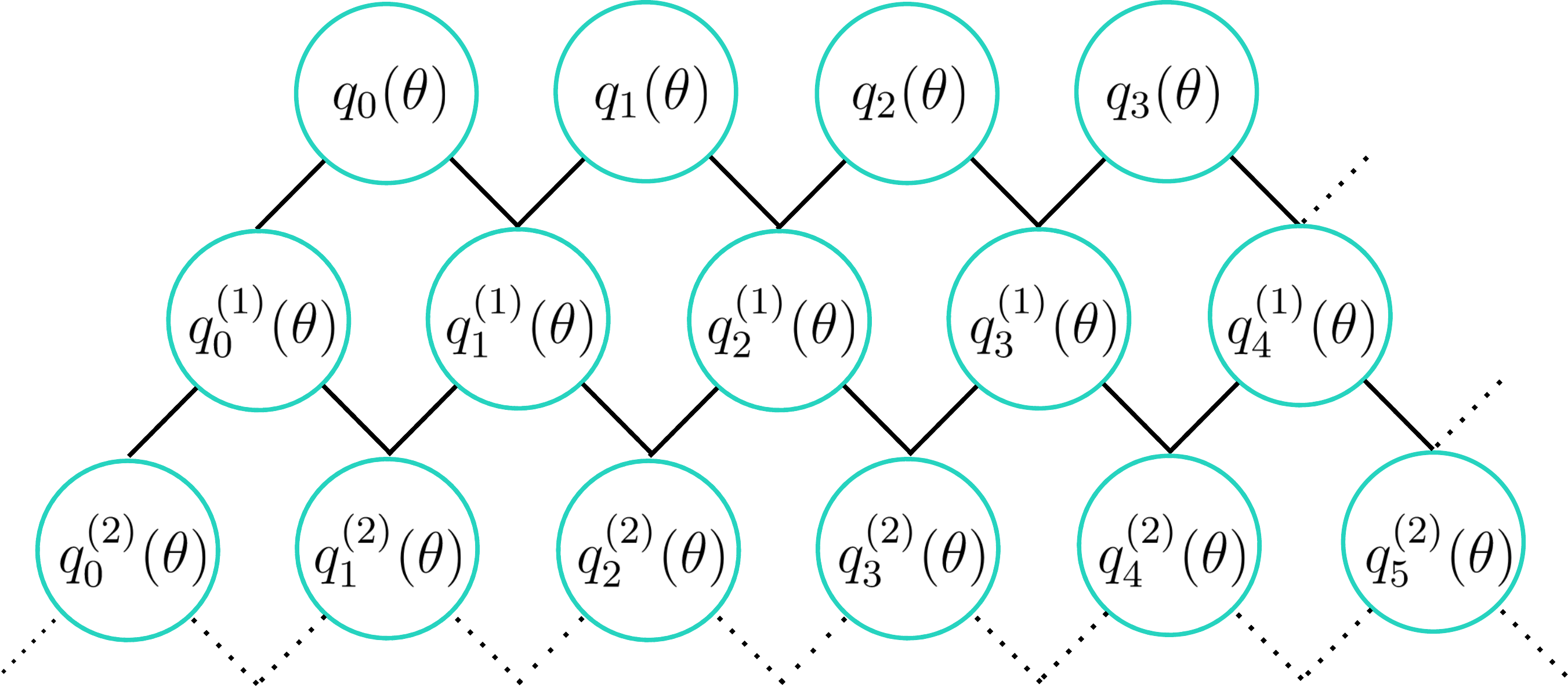}
     \caption{Sketch of the recursive definition of the trigonometric polynomials in Theorem \ref{thm:hybrid_hierarchy} which. The trigonometric polynomials in the $b$'th layer of the ``triangle'' correspond to the polynomials whose nonnegativity is demanded in the $b$'th level of the hierarchy.}
     \label{fig:poly_sketch}
 \end{figure}
\begin{Theorem}
\label{thm:hybrid_hierarchy}
    Let $f(x,y)$ be a  real bivariate polynomial with degree $2m$ that is positive on $\mathbb{R}^2 \setminus \{0\}$. Then there exists an $b \in \mathbb{N}$ such that $q_{s}^{(b)}(\theta)\geq 0$ for all $s=0,\dots,2m+N$ where $q_{s}^{(b)}(\theta)$ are the real univariate trigonometric polynomials defined via the decomposition 
    \begin{equation}
        (1+r)^b f(r,\theta) = \sum_{s=0}^{2m+b} q_{s}^{(b)}(\theta) \, r^s.
    \end{equation}
\end{Theorem}
\textit{Proof:} The negation of the statement leads to a contradiction to P\'olya's theorem. Assume that $f$ is positive on $\mathbb{R}^2 \setminus \{0\}$ and there exists an angle $\widetilde{\theta}\in [0,2\pi]$ such that the sequence $v_{i} := \min_{s \in \{0,\dots,2m+i\}} q_{s}^{(i)}(\widetilde{\theta})$ takes only negative values. Now, one considers the univariate polynomial $w(r):=f(r,\widetilde{\theta})$. By assumption, $w(r)$ is positive on $\mathbb{R}_{+} \setminus \{0\}$ and there exists no $b \in \mathbb{N}$ such that all coefficients of $(1+r)^{b}w(r)$ (given by  $q_{s}^{(b)}(\widetilde{\theta})$) are nonnegative. This contradicts P\'olya's theorem which finishes the proof. $\Box$

Theorem \ref{thm:hybrid_hierarchy} can be applied to obtain converging sequences of upper bounds to the optimisation problems to detect spin nonclassicality~\eqref{eq:spin_classicality_optimisation_problem} and nonclassicality of the harmonic oscillator~\eqref{eq:opt_for_harm_osc}.  

For spin systems, the polynomials under consideration $\widetilde{p}_{V}(\bar{\beta}, \beta)$ are all of the form~\eqref{eq:stereographic_polynomial} which in polar coordinates $\beta=e^{i\theta}r$ transforms as 
\begin{align}
      \widetilde{p}_{V}(\bar{\beta}, \beta)  &= \sum_{k,j=0}^{2m} \widetilde{V}_{kj} \bar{\beta}^{k} \beta^{j} \\
          &= \sum_{k,j=0}^{2m} \widetilde{V}_{kj} e^{i \theta (j-k)} r^{k+j} \\
          &= \sum_{s=0}^{2m} \sum_{k+j=s} \widetilde{V}_{kj} e^{i \theta (j-k)} r^{s} \\
          &= \sum_{s=0}^{2m} q_{s}^{V}(\theta) r^{s}
\end{align}
where $q_{s}^{V}(\theta)$ are the trigonometric polynomials associated to the observable $V$. These are explicitly given by 
\begin{equation}
    q_{s}^{V}(\theta) := \sum_{k+j=s} \widetilde{V}_{kj} e^{i \theta (j-k)}. 
\end{equation}
The hierarchy of SDPs to detect nonclassicality of a state $\rho$ of $m$ qubits can now be written as 
\begin{equation}
\label{eq:polya_fejer_sdp}
     \begin{array}{lll}
          l_{b} = &\underset{V}{\min} &\mean{V}_{\rho}  \\
          & \textnormal{s.t} \; & {q_{s}^{V}}^{(b)}(\theta) \geq 0 \textnormal{ for all } s= 0,\dots,2m+b \\
     \end{array}
\end{equation}
where ${q_{0}^{V}}^{(b)}(\theta) := {q_{0}^{V}}(\theta)$, ${q_{2m+b}^{V}}^{(b)}(\theta) := {q_{2m}^{V}}(\theta)$ and ${q_{s}^{V}}^{(b)}(\theta) := {q_{s-1}^{V}}^{(b-1)}(\theta) + {q_{s}^{V}}^{(b-1)}(\theta)$ for $s=1,\dots,2m+b-1$.
This recursive definition of the trigonometric polynomials ${q_{s}^{V}}^{(N)}(\theta)$ is reminiscent of the definition of Pascal's triangle and is depicted in  the main text in Fig. 3b, which is enlarged in Figure \ref{fig:poly_sketch}. 

By Theorem \ref{thm:pos_trig_polynomial}, the number of variables in the SDP \eqref{eq:polya_fejer_sdp} amounts to $\mathcal{O}(m+b)$ positive semidefinite matrices of size less than $(m+1)\times(m+1)$ leading to $\mathcal{O}(m^{3}+m^{2} b)$ scalar variables. On the other hand, the number of scalar variables for Reznick's hierarchy scales as $\mathcal{O}((m+b)^{4})$.

\section{Finite rays for approximating the classical set from inside} 
\label{app:finite_lines}
Both Reznick's hierarchy and the hierarchy derived in Appendix~\ref{app:hybrid_hierarchy} provide inner approximations to the nonnegative bivariate polynomials. The finite rays approximation as described in the main text gives rise to an outer approximation of this cone.
From the technical perspective, it is slightly more convenient to join rays of opposite directions and consider lines specified by angles $\{\theta_i\}_{i=1}^{N} \subsetneq [0,\pi]$.
Given a bivariate polynomial in $f(r,\theta)$ in polar coordinates and a finite set of lines specified by angles $\{\theta_i\}_{i=1}^{N} \subsetneq [0,\pi]$, the idea is to consider the corresponding set of univariate polynomials $f_{i}(r)$ defined via $f_{i}(r):=f(r,\theta_{i})$. 
It can be easily seen that the nonnegativity of $f$ implies the nonnegativity of all $f_i$.
As a consequence, every subset of $[0,\pi]$ gives rise to an outer approximation of the cone of bivariate nonnegative polynomials.
Applying to the problem of nonclassicality of the harmonic oscillator, this gives rise to the lower bounds $v_{D,T} \leq v_D$ computed by the SDP  
\begin{equation}
     \begin{array}{lll}
        \textnormal{given} &\rho, &T=\{\theta_i\}_{i=1}^{N}\\
          v_{D,T} = &\underset{W}{\min} &\mean{W}_{\rho}  \\
          & \textnormal{s.t} \; & p_{W}(r, \theta_{i}) \geq 0 \textnormal{ for all } i \\
          & \; &\deg(p_{W}) \leq D.
     \end{array}
\end{equation}
Applying to the problem of nonclassicality in spin systems, this gives rise to the the lower bounds $l_{T} \leq l$ computed by the SDP
\begin{equation}
     \begin{array}{lll}
        \textnormal{given} &\rho, &T=\{\theta_i\}_{i=1}^{N}\\
          l_{T} = &\underset{V}{\min} &\mean{V}_{\rho}  \\
          & \textnormal{s.t} \; & \tilde{p}_{V}(r,\theta_{i}) \geq 0 \textnormal{ for all } i. 
     \end{array}
\end{equation}
From a practical point of view the approximation provides a simple, effective method to estimate the optimality gap, and one can expect that one obtains a converging behavior below pre-described error bounds (as seen in earlier examples).

\bibliography{ref}